\documentclass[
    aps,
    pra,
    longbibliography,
    superscriptaddress,amsmath,amssymb,amsfonts, 
    tightenlines,
    twocolumn,
    ]{revtex4-2}

\usepackage[utf8]{inputenc}  
\usepackage[T1]{fontenc}     
\usepackage[english]{babel}  

\usepackage[colorlinks = true, linkcolor=teal, citecolor=teal, urlcolor  = teal]{hyperref}

\usepackage{graphicx} 
\usepackage[babel]{microtype}  
\usepackage{amsmath,amssymb,amsthm,bm,amsfonts,mathrsfs,bbm} 
\usepackage{setspace}
\usepackage{times}
\usepackage{braket}
\usepackage[bottom]{footmisc}
\usepackage{blochsphere}
\usepackage{pgfplots}
\usepackage{amsthm}

\pgfplotsset{compat=1.17}
\usetikzlibrary{calc,3d,shapes, pgfplots.external, intersections}
\usepackage{algorithm}
\usepackage{physics}
\usepackage[noend]{algorithmic}
\usepackage{eqparbox}

\usepackage{tikz}
\usepackage{mathtools}
\usetikzlibrary{arrows.meta}
\usepackage{csquotes}

\usepackage{xspace}  
\usepackage{pgfplots}
\usepackage{verbatim}
\usepackage{amsmath}

\newtheorem{theorem}{Theorem}

\newtheoremstyle{note}
{3pt}
{3pt}
{}
{}
{\itshape}
{:}
{.5em}
{}

\newcommand\id{\leavevmode\hbox{\small1\kern-3.3pt\normalsize1}}

\newcommand{\x}{{\bf x}}

\newtheorem{corollary}[theorem]{Corollary}

\definecolor{jens}{rgb}{0.1,0.4,0.6}

\usepackage{pdfpages}
\usepackage{pgffor}
\makeatletter
\AtBeginDocument{\let\LS@rot\@undefined}
\makeatother

\begin{document}

\title{Learning with errors may remain hard against quantum holographic attacks}

\author{Yunfei Wang}
\email{yunfeizl@umd.edu}
\affiliation{Joint Center for Quantum Information and Computer Science, NIST/University of Maryland, College Park, MD 20742, USA}
\affiliation{Joint Quantum Institute, University of Maryland and NIST, College Park, Maryland 20742, USA}
\affiliation{Department of Computer Science, The University of Pittsburgh, Pittsburgh, PA 15260, USA}


\author{Xin Jin}
\email{XIJ90@pitt.edu}
\affiliation{Department of Computer Science, The University of Pittsburgh, Pittsburgh, PA 15260, USA}

\author{Junyu Liu}
\email{junyuliu@pitt.edu}
\affiliation{Department of Computer Science, The University of Pittsburgh, Pittsburgh, PA 15260, USA}

\maketitle

{\bf The quantum hardness assumption of the Learning with Errors (LWE) problem serves as the foundation of modern cryptography and as a central assumption of lattice-based post-quantum cryptography. Recent advances have shown that estimating entanglement entropy is computationally as hard as the LWE problem, raising a puzzle in the context of the study of quantum gravity and AdS/CFT, where entropies appear to be computable through extremal surface areas. Such a connection might suggest a novel quantum algorithm to attack the LWE problem through constructing holographic duals and performing computationally simple measurements of quantum extremal surfaces. This tension seems to imply one of three possibilities: that AdS/CFT duality itself is computationally intractable, that the quantum-extended Church–Turing thesis (QECTT) is violated, or that LWE is easier than currently believed. However, in this work, we rigorously examine a fourth resolution: the operational task of estimating surface areas to the required precision is itself computationally hard. We formalize this by developing two holographic quantum algorithms. For $\mathcal{O}(N)$ entropy differences, we show that measuring Ryu–Takayanagi (RT) geodesic lengths via heavy-field two-point functions requires exponentially many measurements in $N$, even when the boundary CFT state is efficiently preparable. For $\mathcal{O}(1)$ corrections, we show that reconstructing the bulk covariance matrix and extracting entropy requires $2^{\mathcal{O}(N)}$ time. Although computationally hard, we also compare the efficiency with the Block Korkine-Zolotarev (BKZ) lattice reduction algorithm for LWE. These results help reconcile the apparent tension with QECTT, demonstrating that holographic entropy is consistent with quantum computational limits without assuming a computationally hard holographic dictionary, and providing new insights into the quantum cryptanalysis of lattice-based cryptography.}

\section{{Introduction}}\label{sec:introduction}

Learning With Errors (LWE) is a canonical average-case hard problem underlying modern lattice-based post-quantum cryptography. An LWE instance is specified by a dimension $n$, modulus $q$, and an error distribution (typically a discrete Gaussian): an adversary receives samples ($\mathbf{a}_i, b_i$) with $b_i=\left\langle\mathbf{a}_i, \mathbf{s}\right\rangle+e_i (\bmod q)$ for a secret $\mathbf{s} \in \mathbb{Z}_q^n$ and iid errors $e_i$. The central tasks are either to recover $\mathbf{s}$ (search-LWE) or to distinguish the noisy linear samples from uniform (decision-LWE). LWE admits worst-case/averagecase reductions to lattice problems and forms the hardness backbone for a large suite of cryptographic constructions; its presumed intractability against both classical and quantum adversaries is the foundation of much of post-quantum cryptography \cite{regev2024latticeslearningerrorsrandom}. In 2025, the US government chooses three primary lattice-based cryptography, CRYSTALS-KYBER \cite{bos2018crystals}, CRYSTALS-Dilithium \cite{ducas2018crystals} and FALCON \cite{prest2020falcon} for post-quantum migration through NIST's standarization program \cite{nist}. Those algorithms are based on hard lattice problems that can be reduced to variants of LWE. Thus, the quantum hardness assumptions of LWE serve as foundations of digital security in the future post-quantum era, when a Shor's quantum device is capable to break existing asymmetric cryptographic systems based on prime factoring and elliptic curves \cite{shor1999polynomial}.  

In \cite{gheorghiu2024estimatingentropyshallowcircuit}, Gheorghiu and Hoban introduced the entropy-difference problem, which asks one to decide which of two efficiently preparable quantum states has larger entanglement entropy, given the promise that the two values differ by at least a constant. They studied two natural formulations. In the circuit model (QED$_\delta$, Quantum Entropy Difference), the states are generated by shallow quantum circuits acting on $(N+K)$ qubits, and the entropy is computed on a chosen $N$-qubit subsystem. In the Hamiltonian model (HQED$_\delta$, Hamiltonian Quantum Entropy Difference), the states are ground states of local Hamiltonians, and the entropy is defined on a subsystem of comparable size. In both cases the problem is framed as an operational task of comparing reduced entropies, with explicit bounds on computational resources such as circuit depth, state-preparation cost, and the number of copies or measurements. This careful formulation makes it possible to analyze the complexity of entropy estimation and to connect it directly to cryptographic hardness assumptions. The main result in ref.~\cite{gheorghiu2024estimatingentropyshallowcircuit} is the discovery of a polynomial-time reduction tying the entropy-difference decision problems (QED/HQED) to LWE, proving that the entropy task inherits LWE-level hardness. Informally, this means that an algorithm which could efficiently decide which of two shallowly-preparable (or ground-state) reduced density matrices has larger von Neumann entropy would immediately yield an efficient solver for LWE. 

However, in another seemingly independent area of study, quantum gravity, the so-called AdS/CFT correspondence (Anti De-Sitter/Conformal Field Theory correspondence \cite{Maldacena1997LargeN,Witten1998AdS}) seems to point out a possible shortcut for solving the HQED problem for some special cases \cite{gheorghiu2024estimatingentropyshallowcircuit}. For special cases we mean that the quantum states in the HQED problem can be described as a vacuum state of some CFT Hamiltonian and those CFT Hamiltonians describes a system with a bulk dual. In holographic duality, the entanglement entropy $S_A$ of a boundary region $A$ is related, at leading order in $1/N$, to the area of a minimal or extremal surface $\gamma_A$ in the bulk AdS spacetime through the Ryu-Takayanagi (RT) formula \cite{Ryu_2006}:
\begin{equation}
    S_A = \frac{\text{Area}(\gamma_A)}{4 G_N}~,
\end{equation}
where $G_N$ is Newton's gravitational constant. 

This relationship suggests that measuring entanglement entropy might reduce to the seemingly simpler task of measuring geometric quantities in the bulk, such as areas of surfaces. As a result, this raised the following question question: could one bypass this hardness by constructing the dual AdS state and directly measuring the extremal surface area—potentially solving problems like LWE that are believed to be intractable \cite{afosr}? 

There are three straightforward possibilities to resolve this paradox. Firstly, there are some evidence that holographic duality itself is computationally hard \cite{bouland2019computational,akers2024holographic}. Secondly, contradicting widely held beliefs in cryptography, it might be that LWE is easy. Finally, one could entertain a radical resolution: the quantum-extended Church–Turing thesis (QECTT) might fail. If bulk-area measurements can be performed physically with polynomial resources, yet no quantum algorithm can efficiently reproduce their outcomes — for instance because doing so would yield an efficient solver for LWE — then physical processes would effectively compute beyond BQP, in direct tension with QECTT. \cite{susskind2020horizonsprotectchurchturing}

In this work, we attempt to examine a fourth possibility: the measurement process of the Ryu-Takayanagi geodesic length, being a global quantity, and the fine-grained entropy might be fundamentally hard to resolve. We frame this through the lens of AdS/CFT as an efficient communication channel between two observers: i) a CFT observer who prepares states in the boundary theory; ii) an AdS observer at the conformal boundary who operates in the dual bulk geometry. The AdS/CFT dictionary functions as an efficient channel: when a boundary observer prepares a CFT state, the dual bulk observer effectively finds themselves embedded in the corresponding AdS spacetime, but without knowledge of the geometric details. Note that our bulk observer is always located at the conformal boundary, where all observables are defined unambiguously \cite{Witten2023Algebra,Witten2023Background}. Crucially, to preserve the QECTT, the computational hardness of dual operations must match for both observers. To formalize this, we develop two quantum algorithms based on holographic duality. Our findings can be summarized into the two theorems shown below (assuming that the $N$ qubit state can be prepared efficiently in $\log$-depth circuit). Thus, our work aims to point out a new observation: even while AdS/CFT provides an efficient mapping between states, the intrinsic hardness of entropy measurement is preserved across both descriptions.

\begin{theorem} [Measurement complexity of RT entropy at leading order (informal)] \label{thm:GeoLenMeaComLeadingMain}

    Estimating the Ryu-Takayanagi geodesic length in a holographic CFT state to within $\mathcal{O}(1)$ accuracy already requires exponentially many number of measurements $M$ in the boundary system size $N$:
    \begin{equation}
        M \gtrsim e^{\mathcal{O}\left(N^{1 / \alpha}\right)}~,
    \end{equation}
    where $1 \lesssim \alpha$. Thus, even the leading-order RT prescription is operationally intractable.
\end{theorem}

\begin{theorem} [Measurement complexity of holographic entropy at $\mathcal{O}(1)$ (informal)] \label{thm:expcomplexity}

    When quantum (FLM) corrections are included, extracting the bulk entanglement entropy with $\mathcal{O}(1)$ precision becomes even harder, with measurement complexity scaling as
    \begin{equation}
        \mathcal{T}(N) \sim \operatorname{poly}(N) e^{\mathcal{O}\left(N^{1 / \alpha}\right)} 2^{\mathcal{O}(N)}~,
    \end{equation}
    showing that fine-grained entropy estimation is exponentially costly.
\end{theorem}

\section{{Results}}\label{sec:results}

In this section, we begin with setting up the Hamiltonian Quantum Entropy Difference (HQED) problem and its connection to LWE following ref.~\cite{gheorghiu2024estimatingentropyshallowcircuit}, then formulate it in AdS/CFT where entropy differences correspond to extremal surface area variations. We demonstrate that for $\mathcal{O}(N)$ differences, measuring Ryu-Takayanagi geodesic lengths via Euclidean two-point functions requires exponentially many measurements. Moreover, for $\mathcal{O}(1)$ differences, Faulkner-Lewkowycz-Maldacena (FLM) corrections \cite{Faulkner2013Quantum} necessitate full covariance matrix reconstruction, again demanding exponential resources. This exact parallel between holographic measurement complexity and LWE hardness resolves the entropy paradox while preserving the quantum-extended Church-Turing thesis.

\subsection{{Quantum entropy difference problem and its relation to LWE}}

The LWE problem \cite{regev2024latticeslearningerrorsrandom} can be formulated in terms of the hardness of distinguishing between two functions in an Extended Trapdoor Claw-Free (ETCF) family. Given a public matrix $A \in \mathbb{Z}_q^{M \times N}$, secret vector $s \in \mathbb{Z}_q^N$, and noise terms $e, e'$ drawn from a suitable error distribution (e.g., discrete Gaussian), we define the following pair of functions:
\begin{equation} \label{eq:ETCFfunction}
    \begin{split}
        f(b,x) & = A x + b \cdot u + e \pmod{q}~,\\
        g(b,x) & = A x + b \cdot (A s + e') + e \pmod{q}~,
    \end{split}
\end{equation}
where $b \in \{0,1\}$, $x \in \mathbb{Z}_q^N$, and $u$ is uniformly random in $\mathbb{Z}_q^M$. The function $f$ is injective in $(b,x)$. The function $g$ is \textit{approximately} 2-to-1: for a fixed output $y$, there are typically two distinct inputs $(b,x)$ mapping to $y$.

Distinguishing whether a given sample comes from $f$ or from $g$ is equivalent to solving the LWE problem by the injective invariance of the ETCFs \cite{gheorghiu2024estimatingentropyshallowcircuit}. If $f$ is sampled, the output distribution is almost uniform; if $g$ is sampled, the structure induced by the hidden $s$ and the two-to-one nature introduces collisions.

In the context of quantum information, we can encode these ETCFs into quantum states with $C_1$ and $C_2$ be quantum circuits acting on $N+K$ qubits. This can be done in logarithmic depth circuits since it invlolves only linear-algebraic operations \cite{Mahadev2018Classical,brakerski2021cryptographic}. Define the following $N$-qubit mixed states::
\begin{equation}
    \begin{split}
        \rho_1 & =\operatorname{Tr}_K\left(C_1|00 \ldots 0\rangle\langle 00 \ldots 0| C_1^{\dagger}\right)~, \\ 
        \rho_2 & =\operatorname{Tr}_K\left(C_2|00 \ldots 0\rangle\langle 00 \ldots 0| C_2^{\dagger}\right)~.
    \end{split}
\end{equation}
These states differ in their entanglement structure. In particular, we have 
\begin{equation}
S\left(\rho_1\right) \geq S\left(\rho_2\right)+1 \quad \text { or } \quad S\left(\rho_2\right) \geq S\left(\rho_1\right)+1~,
\end{equation}
promised that one of these is the case. This is known as the \textit{\textbf{quantum entropy difference problem (QED)}}. There is a Hamiltonian analogue of QED which is called the \textit{\textbf{Hamiltonian Quantum Entropy Difference (HQED)}}. The problem amounts to considering \textbf{\textit{ground states}} of $N+K$ qubits: $\ket{\psi_1}$ and $\ket{\psi_2}$ of two local Hamiltonians $H_1$ and $H_2$ respectively. The two $N$-qubit mixed states are defined as 
\begin{equation}
\begin{split}
\rho_1 = \operatorname{Tr}_K \left(\left|\psi_1\right\rangle \left\langle\psi_1\right|\right)~, \quad \rho_2 = \operatorname{Tr}_K \left(\left|\psi_2\right\rangle \left\langle\psi_2\right|\right)~.
\end{split}
\end{equation}
Both the QED problem and HQED problem was proven to be in the same complexity class with the LWE problem \cite{gheorghiu2024estimatingentropyshallowcircuit}. Note that to amplify the entropy gap between the two states, one can generalize the above construction by redefining the ETCF family to distinguish between an injective function and a $2^k$-to-1 function, thereby enlarging the entropy separation to $k$ bits. Hence, efficiently solving the entropy difference problem, even just to estimate the entropy efficeintly at $\mathcal{O}(N)$, implies an efficient quantum attack on LWE. For more systematic setup, we refer to ref.~\cite{gheorghiu2024estimatingentropyshallowcircuit} and the Supplementary Material.

\subsection{{The entropy difference problem in holographic duality}} \label{subsec:HQEDinHolography}

A related thought experiment, first proposed in ref.~\cite{gheorghiu2024estimatingentropyshallowcircuit}, is based on the following assumptions. First, CFT states can be efficiently prepared from classical descriptions of functions $f$ (1-to-1) and $g$ (2-to-1) (eg., ETCF functions from Eqn.~(\ref{eq:ETCFfunction})). Moreover, these functions yield boundary states with a constant, detectable entropy difference. We will comment on these assumptions later.

\begin{figure}[!ht] 
\centering
\includegraphics[width=0.25\textwidth]{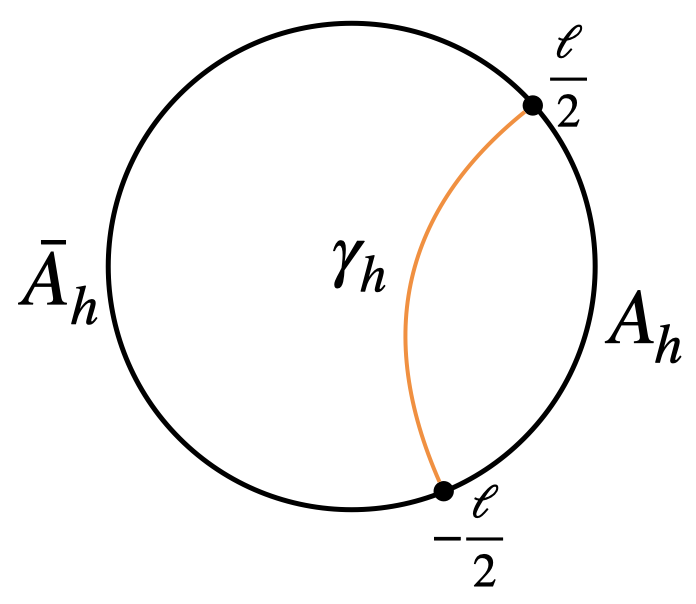}
\caption{Boundary part $A_h$ corresponding to the CFT state $\ket{\psi_h}$ and its complementary part $\bar{A}_h$.}
\centering \label{Fig:BoundaryBulkState}
\end{figure}

The protocol is as follows. Given a function $h$ promised to be either $f$ or $g$, Alice prepares the CFT ground state $|\psi_h\rangle_{\text{CFT}}$ and consider its subregion entropy for some fixed boundary condition $\ell$. If $h = f$, the entropy is higher; if $h = g$, the entropy is lower. Via AdS/CFT, the state has a bulk dual $|\psi_h\rangle_{\text{bulk}}$, with entropy related to the area of a minimal surface $\gamma_h$ through the RT formula (see Figure.~\ref{Fig:BoundaryBulkState}).

A bulk observer measures the area of $\gamma_h$ and if the measured area exceeds a threshold, he drastically change the geometry to make it corresponds to a state as orthogonal as possible to the original state. He will do nothing if the measured area does not exceeds a threshold. This physical process that the geometry evolves is described by $H_{\text{bulk}}$. Translating this back to the boundary, Alice obtained $H_{\text{CFT}}$, then performs a SWAP test. If her final state is unchanged, she infers $h = f$; otherwise, $h=g$.

An efficient implementation of this protocol would solve a specialized LWE variant by distinguishing injective ($1$-to-$1$) from degenerate ($2^{a N}$-to-$1$, $a N \in \mathbb{Z}^+$, $a<1$) ETCF functions, contradicting the widely believed LWE hardness. 

However, the above construction being an efficient algorithm depends on the assumption that the bulk observer can efficiently measure the RT surface area. To analyze the operational difficulty of estimating entanglement entropy in AdS/CFT, we begin with the semiclassical Ryu-Takayanagi prescription in AdS$_3$/CFT$_2$. The central charge $c$ of the boundary CFT, which characterizes the number of degrees of freedom, is related to the bulk Newton constant $G_N$ and AdS radius $R$ via the Brown-Henneaux formula \cite{BrownCentral1986}: $c=\frac{3 R}{2 G_N} \sim N~,$ where $N$ represents the number of boundary degrees of freedom or, equivalently, the number of qubits in a discretized version of the CFT.

This relation implies that the bulk Newton constant scales inversely with $N$, i.e., $G_N \sim 1 / N$, and likewise the Planck length in $2+1$ dimensions scales as $\ell_{\mathrm{Pl}} \sim \mathcal{O}(1 / N)$. Applying the RT formula to compute the entanglement entropy of an interval of length $\ell$ on the boundary yields:
\begin{equation}
S=\frac{L_{\mathrm{RT}}}{4 G_N} \sim \mathcal{O}(N) \quad \Rightarrow \quad L_{\mathrm{RT}} \sim \mathcal{O}(1)~,
\end{equation}
which indicates that estimating the leading-order entropy thus requires measuring $L_{\mathrm{RT}}$ to within order-one precision, a task that remains within the domain of semiclassical geometry and is, in principle, accessible using classical probes, e.g. heavy mass field two point functions. Note that whether the measurement protocol can be efficiently implemented remains a question which we will explore in the following subsection. 

However, resolving subleading corrections—on the order of a single bit—demands measuring $L_{\mathrm{RT}}$ with precision $\delta L_{\mathrm{RT}} \sim \mathcal{O}(1 / N)$, matching the Planck length in three dimensions. This delineates a transition in measurement complexity: from classical to quantum gravitational sensitivity.

To make this concrete, consider the analytic form of the RT geodesic length in AdS$_3$:
\begin{equation}
L_{\mathrm{RT}}=2 R \log \left(\frac{\ell}{\epsilon}\right)~.
\end{equation}
Here, both $\ell$ (the boundary subregion length) and $\epsilon$ (the UV cutoff) are fixed by boundary conditions. Hence, any fine-grained variation in $L_{\mathrm{RT}}$ must arise from variation in the AdS radius $R$. Resolving such differences to $\mathcal{O}(1 / N)$ precision therefore entails distinguishing between bulk geometries with slightly differing curvature radii. It's also important to realize that due to the fact that since the only varying parameter in $L_{\mathrm{RT}}$ is $R$, then this is, in fact, a global quantity that we are estimating, which is what makes this measurement nontrivial even to leading order. 

In this sense, the requirement of Planck-scale precision is not simply a limitation of the measurement apparatus—it reflects the intrinsic quantum nature of the spacetime itself. The inaccessibility of subleading entropy corrections is thus a manifestation of quantum gravitational effects, rooted not in bulk excitations, but in the quantization of the background geometry. A more detailed analysis of these effects will be presented in the Supplementary Materials.

\subsection{{Measurement complexity of the RT geodesic}} \label{subsec:MeaComRTGeo}

As discussed, estimating the Ryu–Takayanagi (RT) geodesic length to $\mathcal{O}(1)$ precision suffices to compute the leading-order entanglement entropy in the AdS bulk. Operationally, this can be achieved by measuring the bulk two-point function $\langle\phi(X_1)\phi(X_2)\rangle$ of a heavy scalar field in a fixed background. We now quantify the resource cost of such measurements using quantum estimation theory \cite{Helstrom1969Quantum,paris2004quantum}.

Let $O$ denote the some quantum observable, independently measured over $M$ identical copies. The variance of the sample mean $\bar{O}$ is:
\begin{equation}
\operatorname{Var}(\bar{O})=\frac{\sigma_O^2}{M}, \quad \Rightarrow \quad \delta O=\frac{\sigma_O}{\sqrt{M}}~.
\end{equation}
This is the \emph{standard quantum limit} (SQL), describing the noise floor of independent measurements. By contrast, using entangled probe states and collective (non-independent) measurements one can in principle reach the Heisenberg limit, $\delta O_{\mathrm{Heis}} \sim 1/M$ which is a quadratic improvement over the SQL (not exponential). Achieving this scaling requires highly entangled inputs and collective readout and is extremely fragile to noise and imperfect state preparation \cite{Giovannetti2004,Giovannetti2011}; accordingly, the availability of Heisenberg scaling will not remove the exponential-in-$N$ operational obstructions we identify in the holographic setting as shown below .

In the leading order WKB approximation, which is valid only for heavy mass fields, the two-point function decays exponentially with geodesic distance:
\begin{equation}
\left\langle\phi\left(X_1\right) \phi\left(X_2\right)\right\rangle \sim e^{-m L_{\mathrm{gco}}\left(X_1, X_2\right)}~.
\end{equation}
To resolve $\delta L_{\text{geo}} \sim \mathcal{O}(1)$, one must detect a signal difference as the following, hence implies the number of measurements $M$:
\begin{equation} \label{eq:MeasureGeodesicCom}
\delta\left\langle\phi\left(X_1\right) \phi\left(X_2\right)\right\rangle \sim e^{-m \delta L_{\mathrm{gco}}} \Rightarrow M \gtrsim e^{2 m \delta L_{\text{geo}}}~.
\end{equation}

Moreover, the probe's backreaction must remain negligible. The perturbation to the geometry from a massive field scales as: $\delta L \sim G_N m \sim \frac{m}{N} \ll 1~,$ where $G_N \sim 1/N$ in AdS units. To suppress backreaction, we require $m \ll N~.$ Note that we assumed that $N$, the number of qubits, to be large, but still efficiently prepared. We only consider runtime exponential in $N$ to be inefficient. 

For starters, it is natural to use a moderately heavy probe with $m \sim \sqrt{N}$, the WKB approximation remains valid, ensuring that the boundary two-point correlator retains its standard geometric interpretation in terms of bulk geodesic lengths. However, the measurement cost derived from Eqn.(\ref{eq:MeasureGeodesicCom}) scales as $M \gtrsim e^{\mathcal{O}(N)},$ so even the leading-order estimation of entropy demands exponentially many samples. This establishes that in the heavy-probe regime, the geodesic method cannot circumvent the exponential bottleneck. One might ask whether advanced measurement-compression techniques, such as classical shadows \cite{Huang_2020,Wang:2024ygz}, could reduce this cost. Classical shadow tomography allows one to reconstruct expectation values of many observables simultaneously from a small number of randomized measurements. For an operator $\mathcal{O}$ acting on an $N$-quibit system, the typical sample complexity scales as
\begin{equation}
    M_{\text {shadow }} \sim \frac{\log K}{\epsilon^2} \operatorname{Tr}\left(\rho \mathcal{O}^2\right)~,
\end{equation}
where $K$ is the number of observables to be estimated and $\epsilon$ the target additive error. When the observable is few-body or has bounded operator norm, this scaling can be polynomial in $N$. However, we are required to measure this two-point function to an exponential preciseness Eqn.~(\ref{eq:MeasureGeodesicCom}). Since $M \sim \epsilon^{-2}$, this sample complexity is still exponential in $N$. Even viewing this problem more generally, to estimate an extensive entropy difference of order $\alpha N$ for an $N$ qubit system, classical-shadow tomography offers no exponential advantage for holographic ground states. Although such states are ground states of local boundary Hamiltonians, their entanglement structure is intrinsically nonlocal: the entropy is encoded in global correlations corresponding to bulk geometric data. Shadow tomography is efficient only for few-body or bounded-norm observables, whereas the operators relevant for holographic entropy estimation have exponentially large variance. Consequently, the sample complexity for resolving an $\mathcal{O}(N)$ entropy gap remains exponential in $N$; the exponential hardness originates from the holographic encoding itself rather than from measurement inefficiency.

On the other hand, for a lighter probe with $m \sim \log N'$, one might hope for a substantial reduction in measurement cost, since in this case $M$ scales only as $\operatorname{poly}(N')$. Yet the WKB condition $m L_{\text{geo}} \gg 1$ imposes a severe consistency requirement: achieving semiclassical validity forces $\log N' \sim \sqrt{N}$, hence $N' \sim e^{\sqrt{N}}$. The boundary Hilbert space must therefore be exponentially enlarged relative to the heavy-mass case. Preparing the corresponding state already incurs exponential overhead. So the apparent polynomial measurement savings are effectively illusory. In other words, shifting to the light-probe regime only relocates the exponential difficulty—from measurements to state preparation—without eliminating it.

These observation are then summarized in to Thm.~\ref{thm:GeoLenMeaComLeadingMain} stated before. Thus, under physically reasonable conditions, the measurement of the RT geodesic length to $\mathcal{O}(1)$ accuracy is not efficient. This means that even leading-order entropy in holographic settings is generically computationally intractable. This result matches the expected exponential hardness for LWE problem with ETCF functions consists of a $2^{aN}$-to-1 function and a 1-to-1 function or equivalently the HQED problems where the entropy gap is $\sim a N$, with $a \lesssim 1$.


\subsection{{Measurement Complexity of Quantum Corrections}} \label{subsec:MeaComQuantumCorr}

We now turn to the problem of estimating the entropy of the ground state of a local physical Hamiltonian up to $\mathcal{O}(1)$ precision. This level of resolution is physically motivated by the LWE, QED, and HQED problems, which all require distinguishing entropy differences of $\mathcal{O}(1)$. At such scales, semiclassical approximations break down, and the RT formula must be supplemented by quantum corrections. These are captured by the Faulkner–Lewkowycz–Maldacena (FLM) prescription \cite{Faulkner2013Quantum}, which adds bulk entanglement entropy to the geometric RT term:
\begin{equation} \label{eq:FLMCorrections}
\begin{aligned}
S(A) &= S_{\text{cl}}(A) + S_q(A) + \mathcal{O}(G_N), \\
S_q &=S_{\text{bulk-ent}} + S_{\ldots}\\
& = S_{\text{bulk-ent}}+\frac{\delta A}{4 G_N}+\left\langle\Delta S_{\text{Wald}-\text{like}}\right\rangle+S_{\text {counterterms }}
\end{aligned}
\end{equation}
These FLM correction are consists of $S_{\text{bulk-ent}}$ which is the bulk entanglement entropy across the RT surface, and a integral of local geometric term, which we collectively denote by $S_{\ldots}$. Hence, to estimate the holographic entanglement entropy to $\mathcal{O}(1)$, beside the measurement of the RT geodesic length described in previous section, we also need to estimate all quantities: $S_{\text{bulk-ent}} + S_{\ldots}$ in the FLM correction.



Operationally, it is the bulk entanglement term $S_{\text {bulk-ent}}$ that poses the principal challenge, as it requires reconstructing fine-grained quantum information across the RT surface. Hence, let's start by attempting the measurement of $S_{\ldots}$. 

The Wald-like contributions $\Delta S_{\text{Wald-like}}$ arise from higher-derivative corrections to the bulk action, generated for instance by integrating out heavy fields coupled to gravity \cite{Dong2014HigherDerivative,Cheung:2018cwt}. These terms, together with local counterterms $S_{\text{counterterms}}$, enter the entropy via generalized Wald formulas and can be operationally extracted by probing how heavy bulk fields renormalize local gravitational couplings. In practice, both contributions imprint themselves on boundary correlators through the same geodesic X–ray kernel; hence their inference inherits the exponential measurement complexity of geodesic length reconstruction (Thm.~\ref{thm:GeoLenMeaComLeadingMain}).

Moreover, the geometric backreaction term $\delta A/4G_N$ is determined by the linear response of the bulk metric to $\langle T_{\mu\nu}\rangle$, the expectation value of the bulk field stress tensor. Here the $G_N$ dependence cancels, leaving an $\mathcal{O}(1)$ linear functional of the stress tensor with a smooth kernel. A bulk-local observer could estimate this contribution by measuring this one point function with only polynomial resources using standard quantum-limit scaling. For more details on this part, we refer to the Supplementary Materials. Putting these together, we obtain the following corollary.

\begin{corollary} [Complexity for estimating $S_{\ldots}$ (informal)] \label{cor:Scl+S...Complexity}

Extracting $S_{\text{cl}}$ together with the Wald-like, counterterm, and $\delta A/4G_N$ corrections requires overall exponential measurement cost in $N$, dominated by the reconstruction of geodesic lengths, with only polynomial overhead from the additional local terms.
\begin{equation} 
\mathcal{T}(N) \sim \operatorname{poly}(N) e^{N^{1/\alpha}}~, 
\end{equation} 
for some $1 \lesssim \alpha$, with the exponential factor dominating the asymptotics.
\end{corollary}


At last, we turn to the estimation of $S_{\text{bulk-ent}}$. To compute the FLM quantum correction, we focus on the leading 1-loop contribution from free bulk fields propagating on a fixed classical background. In this regime, the bulk state is Gaussian and fully determined by its two-point function. Discretizing the entanglement wedge $\mathcal{E}_A$ into $D \sim 2^N$ sites-matching the Hilbert space dimension $\operatorname{dim} \mathcal{H}_A=2^N$ of the boundary subregion-yields $2 D$ canonical degrees of freedom $\left\{\xi_i\right\}=\left\{\phi_i, \pi_i\right\}$.
The full entanglement structure is encoded in the covariance matrix:
\begin{equation}
\Gamma_{i j}=\left\langle\xi_i \xi_j+\xi_j \xi_i\right\rangle-2\left\langle\xi_i\right\rangle\left\langle\xi_j\right\rangle~,
\end{equation}
a real, symmetric $2 D \times 2 D$ matrix. Its symplectic eigenvalues $\left\{\nu_i\right\} \geq 1 / 2$ determine the entropy of each mode, and the total bulk entanglement entropy is given by a sum: $S_{\text {bulk-ent }}=\sum_i S\left(\nu_i\right),$ where $S(\nu)$ is the von Neumann entropy of a single Gaussian mode.

Crucially, resolving generic $S_{\text {bulk-ent}}$ requires access to all symplectic modes in $\mathcal{E}_A$, not just those near the RT surface. Even small deviations in $\nu_i-1 / 2$ must be retained, as they can collectively yield an $\mathcal{O}(1)$ correction. Thus, estimating the FLM term requires reconstructing an exponentially large number of bulk degrees of freedom $\sim 2^N$ from the boundary-a task that is generically computationally intractable, which is in agreement with Thm.~\ref{thm:expcomplexity}. (See the Supplementary Material for details on this)

For free (Gaussian) bulk fields, correlators decay polynomially with distance,
\begin{equation}
    \Gamma_{i j} \sim \frac{1}{\left|x_i-x_j\right|^{\Delta}}, \quad \Delta>1~,
\end{equation}
so $\Gamma$ can be approximated as banded or sparse. This structure permits the use of quantum algorithms for sparse Hamiltonian simulation. In particular, encoding the modular operator $M=i\Omega^{-1}\Gamma$ as a Hamiltonian allows quantum phase estimation (QPE) to estimate symplectic eigenvalues with cost
\begin{equation}
    \operatorname{Cost}_{\mathrm{QPE}}=\mathcal{O}\left(\frac{s}{\varepsilon} \operatorname{polylog}\left(2^{2 N}\right)\right) \sim \mathcal{O}\left(\frac{1}{\varepsilon} \operatorname{poly}(N)\right)~,
\end{equation}
where $s=\operatorname{poly}(N)$ is the sparsity and $\varepsilon$ is the QPE error. Truncating to the $J$ dominant modes ($J$ is at least polynomial in $N$, depending on the details of specific states), one approximates the entropy as
\begin{equation}
    S_{\text {bulk-ent}} \approx \sum_{j=1}^J S\left(\nu_j\right), \quad\left|S-S_{\text {approx}}\right|<\epsilon~.
\end{equation}

The improved scaling is therefore
\begin{equation}
    \text{Build } \Gamma:~ 2^{\mathcal{O}(N)}, \quad \text{Diagonalize: } \frac{J}{\varepsilon} \operatorname{poly}(N), \quad \text{Entropy: $J$.}
\end{equation}
While diagonalization and entropy evaluation can be made efficient, the exponential cost of constructing $\Gamma$ remains unavoidable, since one must still measure $\sim 2^{N}$ correlation functions even in a sparse approximation such that one can decide which $J$ modes are dominant. Together with corollary~\ref{cor:Scl+S...Complexity}, this demonstrates the result presented in Thm.~\ref{thm:expcomplexity}. An alternative way is again to make use of shadow tomography. However, to estimate the fine grained quantum entropy to order $1$ for an $N$ qubit system would require reconstructing eigenvalues of the density operator with exponential accuracy in the Hilbert-space dimension $2^N$. Classical-shadow tomography can efficiently estimate expectation values of many observables, but not the full spectral data of the density matrix unless exponentially many measurements are taken.

This intractability mirrors the hardness of estimating entropy in boundary theories. The HQED and LWE problems demonstrate that computing the von Neumann entropy of quantum states (including ground states of physical local Hamiltonians) to constant additive error is \textit{quantumly hard} under standard cryptographic assumptions. Our results show that the bulk dual physical processes reproduce these boundary complexities from bulk dynamics, affirming that the QECTT is not violated.

\subsection{Complexity comparison: BKZ algorithm vs holography}

It is instructive to compare our holographic algorithm with the best-known algorithms for attacking the LWE problem. A dominant classical approach is to embed the LWE samples into a lattice and use Block Korkine-Zolotarev (BKZ) lattice reduction to recover the secret via closest-vector decoding \cite{korkine1877formes,regev2024latticeslearningerrorsrandom,Laarhoven2016}. Heuristically, after a $\mathrm{BKZ}_\beta$ reduction, the shortest vector length scales as $\lambda_1(\mathcal{L}) \approx \delta_\beta^N(\operatorname{det} \mathcal{L})^{1 / N},$ and decoding succeeds when the error vector is smaller than $\lambda_1$, giving a required blocksize $\beta \sim N$ in the typical cryptographic regime. Solving SVP in this block dominates the runtime, which scales exponentially in $N$,
\begin{equation}
    T_{\mathrm{LWE}} \sim 2^{\mathcal{O}(N)}~.
\end{equation}
Quantum lattice sieving provides only modest constant-factor improvements \cite{Martincryptoeprint,Laarhoven2016}, leaving the exponential scaling intact. Dual and BKW-style combinatorial attacks can slightly reduce the effective exponent for special secret distributions \cite{Guoasiacrypt,Guocryptoeprint}, but do not change the asymptotic picture.

By comparison, our main result (Thm.~\ref{thm:expcomplexity}) shows that extracting the bulk entanglement entropy or related Wald-like corrections in holographic theories requires a measurement cost
\begin{equation}
\mathcal{T}(N)=\operatorname{poly}(N) e^{N^{1 / \alpha}} 2^{\mathcal{O}(N)}=\exp (\mathcal{O}(N))~, \quad 1 \lesssim \alpha~,
\end{equation}
exponential in the number of boundary degrees of freedom. Thus, in a coarse asymptotic sense, both LWE attacks and holographic reconstruction lie in the same hardness class. The difference lies in the resource type: LWE hardness is primarily computational (time/memory), whereas holographic hardness is informational (number of high-precision measurements and covariance entries). Practical distinctions also arise from leading constants in the exponent and the structure of the problem—for example, sparsity of the covariance matrix and dominant symplectic eigenmodes can influence the efficiency of quantum algorithms, but do not change the exponential scaling.

\section{{Discussion}}\label{sec:concoutlook}

Our results show that estimating entanglement entropy in holographic settings—whether at leading order via the Ryu–Takayanagi (RT) formula or through quantum corrections via the FLM prescription—is generically computationally intractable. Even when assuming efficient state preparation and semiclassical bulk dynamics, both measurement and post-processing steps incur exponential cost in the number of boundary degrees of freedom.

This intractability is not a defect but a feature: it ensures consistency between holographic duality and known quantum complexity-theoretic bounds. In particular, entropy estimation remains hard for ground states of physical local Hamiltonians, as formalized by the HQED and LWE problems. Holography faithfully reproduces this difficulty, even in highly symmetric large-$N$ CFTs.

From a bulk perspective, the obstruction arises from operational constraints. Leading-order RT lengths become inaccessible due to precision limits. Whereas extracting order-one FLM corrections requires reconstructing the full covariance structure of $2^N$ bulk modes—an exponentially hard task.

These findings highlight a key aspect of holographic duality: while geometric and entropic observables admit elegant closed-form expressions, their operational extraction from quantum data remains fundamentally hard. This reinforces the validity of the quantum-extended Church–Turing thesis, which asserts that all reasonable physical process—including those described by quantum gravity—can be efficiently simulated by quantum computers. The inaccessibility of entanglement entropy at polynomial cost, even in semiclassical AdS/CFT, underscores a deep interplay between geometry, quantum information, and computational complexity—one that persists across both classical and quantum gravitational regimes.

Recent quantum pseudoentanglement results \cite{aaronson2023quantumpseudoentanglement,bouland2023publickey} which strengthen prior $\mathcal{O}(1)$ entropy-gap claims to $\mathcal{O}(\log{N})$, together with complementary LWE-based constructions, further underscore the computational indistinguishability of low- and high-entanglement states. These results reinforce the general picture that distinguishing entanglement structure can be computationally intractable. But they do not affect our bulk-measurement hardness, which are parametrically larger than the polylogarithmic entropic gaps exhibited in the pseudoentanglement constructions.

Our framework invites broader investigation into the operational complexity of measuring bulk observables in AdS/CFT. For example, estimating thermal entropy from boundary data—such as computing entropy density at finite temperature—appears simple from bulk geometry (e.g., via black hole horizon area), yet remains computationally intractable in practice. Similar challenges arise for transport coefficients like conductivity. These cases exemplify a deeper theme: the strong/weak nature of holography renders many seemingly simple bulk quantities difficult to extract from boundary data, pointing to a rich structure of complexity across the duality.

\section*{{Acknowledgements}}\label{sec:ack}
We thank Hrant Gharibyan, Andru Gheorghiu, Sepher Nezami, John Preskill, and Eugene Tang for numerous insights around 2020. We thank Jens Eisert, Bill Fefferman, and Hsin-Yuan Huang for discussions. YW, XJ and JL are supported in part by the University of Pittsburgh, School of Computing and Information, Department of Computer Science, Pitt Cyber, PQI Community Collaboration Awards, John C. Mascaro Faculty Scholar in Sustainability, NASA under award number 80NSSC25M7057, and Fluor Marine Propulsion LLC (U.S. Naval Nuclear Laboratory) under award number 140449-R08. This research used resources of the Oak Ridge Leadership Computing Facility, which is a DOE Office of Science User Facility supported under Contract DE-AC05-00OR22725.

\bibliographystyle{apsrev4-1}
\bibliography{cite}

\pagebreak
\clearpage
\foreach \x in {1,...,\the\pdflastximagepages}
{
	\clearpage
	

\section*{Supplementary material}

# Supplementary material

## I. LEARNING WITH ERROR AND THE ENTROPY DIFFERENCE PROBLEM

The connection between the quantum information problem and the quantum gravity problem stems from the paper by Gheorghiu and Hoban [1]. We start by reviewing their set up and theorems and end this section with a comments on connections between the entropy problem with AdS/CFT.

Suppose we have a shallow circuit, by shallow we mean given $N$ qubits (or classical random bits) the circuit depth $d$ is $d = \mathcal{O}(\log N)$.

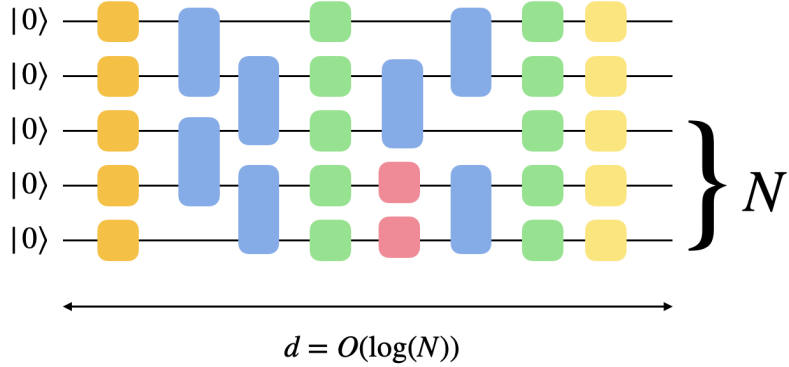

Figure 1. A demonstration of a log depth circuit

The main result of ref. [1] is that to estimate the entropy difference of the result produced by different circuits like this is at least as hard as Learning With Error (LWE). This is considered a "hard" problem, where for "hard" we mean that the best (quantum) algorithm we know requires *exponential* time in $N$. Now we formulate this in a rigorous way. We begin by defining the entropy difference problem and the quantum entropy difference problem, for circuits with depth scaling as $\delta(N)$, for inputs of size $N$ and where $\delta: \mathbb{N} \to \mathbb{N}$, is a monotonically increasing function. First the classical entropy difference problem is defined as the following.

**Definition 1** *($ED_\delta$)*
*Let $C_1$ and $C_2$ be reversible boolean circuits acting on $N + K$ bits such that $\text{depth}(C_1) \leq \delta(N + K)$, $\text{depth}(C_2) \leq \delta(N + K)$. Let*

$$\mathcal{D}_1(y) = \Pr_{x \leftarrow U\{0,1\}^{N+K}} [y = \text{Tr}_K(C_1(x))] \quad \mathcal{D}_2(y) = \Pr_{x \leftarrow U\{0,1\}^{N+K}} [y = \text{Tr}_K(C_2(x))] \ , \tag{1}$$

*denote the output distributions of the circuits when restricted to the first $N$ output bits and with the input chosen uniformly at random. Given $N$, $K$ and descriptions of $C_1$ and $C_2$ as input, decide whether:*

$$S(\mathcal{D}_1) \geq S(\mathcal{D}_2) + 1 \quad \text{or} \quad S(\mathcal{D}_2) \geq S(\mathcal{D}_1) + 1 \ , \tag{2}$$

*promised that one of these is the case.*

Then we also have a similar definition for quantum entropy difference.

**Definition 2** *($QED_\delta$).*
*Let $C_1$ and $C_2$ be quantum circuits acting on $N+K$ qubits such that $\text{depth}(C_1) \leq \delta(N+K)$, $\text{depth}(C_2) \leq \delta(N+K)$. Define the following n-qubit mixed states:*

$$\rho_1 = \text{Tr}_k\left(C_1|00\ldots0\rangle\langle 00\ldots 0|C_1^\dagger\right) \quad \rho_2 = \text{Tr}_k\left(C_2|00\ldots0\rangle\langle 00\ldots 0|C_2^\dagger\right) \ . \tag{3}$$

*Given $N$, $K$ and descriptions of $C_1$ and $C_2$ as input, decide whether:*

$$S(\rho_1) \geq S(\rho_2) + 1 \quad or \quad S(\rho_2) \geq S(\rho_1) + 1 , \qquad (4)$$

*promised that one of these is the case.*

Note that if $\delta(N) = \mathcal{O}(\log(N))$ we obtain the definitions for $\text{ED}_{\log}$ and $\text{QED}_{\log}$, respectively. One can also, in the same way define entropy difference problem for other depth of the circuit. But as proven in ref. [1], the entropy difference problem is easier in shallow circuits. The major theorem proved in [1] is the following:

**Theorem 3** LWE $\leq_P \text{ED}_{\log} \leq_P \text{QED}_{\log}$ .

In here, we summarize the prove and refer the readers who are interested in the detailed prove to ref. [1]. Also note that we restrict our attention to shallow circuits with unbounded fan-out gates. For bounded fan-out case, the understanding is a bit less clear compared to the unbounded fan-out case. The classical entropy difference problem for this case is proven to be easy. The quantum case is not easy, but we also don't know whether it is hard.

First, we state that there exists a reduction from LWE to a problem of distinguishing the two functions in an ETCF ("Extended Trapdoor Claw-Free") family which consists of an 1-to-1 (or injective) function and a 2-to-1 (or claw-free) function. For precise definitions, we refer to [2, 3]. At high level let us suppose that there are two functions, $f: \{0,1\}^N \to \{0,1\}^N$ which is 1-to-1 and $g: \{0,1\}^N \to \{0,1\}^N$ which is *approximately* 2-to-1. One can come up with quantum circuits that generates these states. So for function $f$, we have some quantum circuit $C_f$, such that the output state is

$$\sum_{x \in \{0,1\}^N} |x\rangle |f(x)\rangle , \qquad (5)$$

and for function $g$, we have $C_g$:

$$\sum_{x \in \{0,1\}^N} |x\rangle |g(x)\rangle . \qquad (6)$$

Then we can trace out the register qubit $|x\rangle$ and compute the entropy of the mixed state, for $|f(x)\rangle$, we would have $S(\rho_f) = N$, because it is 1-to-1. And for $|g(x)\rangle$, we would have $S(\rho_g) = N - 1$, because of the collisions that occur while the mapping by the function $g$.

The problem can now be phrased in the following way. Suppose we have some circuit $C_h$, and we are promised that this circuit will create some state:

$$\sum_{x \in \{0,1\}^n} |x\rangle |h(x)\rangle , \qquad (7)$$

and the problem here is to determine whether this function $h$ is a 1-to-1 function or a 2-to-1 function, promised that $h$ is one of these two functions. If we are able to efficiently estimate the entropy of this state with some quantum algorithm, then we would be able to answer this question.

Let's look at the ETCF functions used in the proof more closely. They are defined as:

$$f(b,x) = Ax + b \cdot u + e (\text{mod} q) \quad g(b,x) = Ax + b \cdot (As + e') + e (\text{mod} q) , \qquad (8)$$

with $q \geq 2, b \in \{0,1\}, x \in \mathbb{Z}_q^N, s \in \mathbb{Z}_q^N, A \in \mathbb{Z}_q^{M \times N}, u, e' \in \mathbb{Z}_q^M, e \leftarrow_{D_{\mathbb{Z}_q,B}^M} \mathbb{Z}^M$. The only difference of these two functions is, depending on the value of $b$, we may or may not add $u$ or we may or may not add $As + e'$. This $u$ vector can be thought of as a uniformly random vector of components modulo $q$, whereas $e$ and $e'$ are called the error vectors which will have very small norm, in particular: $||e||_2, ||e'||_2 \ll ||u||_2$.

Then we see that $As + e'$ is very similar to the image of the matrix $A$: $\text{Im}(A)$, whereas if we conditionally add the uniformly random $u$ vector, in general it will not be close to $\text{Im}(A)$. This means that the output of the $f$, conditioned on whether $b$ is 0 or 1, we should have a result that is close to $\text{Im}(A)$ or not close to $\text{Im}(A)$. But for $g$, we would always get an overall output result close to $\text{Im}(A)$. Since the output of $g$ is always close to $\text{Im}(A)$, it's very probable that collisions will happen and $g$ works like the 2-to-1 function whereas $f$ is likely to be injective. Furthermore, the LWE problem is exactly the problem to distinguish between $(A, u)$ and $(A, As + e)$ given the matrix $A$. So determining which is the injective function is as hard as LWE. Moreover, one can show that this can be implemented in log depth

circuits since these functions only involve linear-algebraic operations which is easy to parallel. Moreover, we can consider the following circuits:

$$\begin{aligned}
C_f(b_1, b_2, x, e_u) &= Ax + b_1 \cdot u_1 + b_2 \cdot u_2 + e_u (\mathrm{mod} q) \, , \\
C_{g^{(2)}}(b_1, b_2, x, e_u) &= Ax + b_1 \cdot (As_1 + e'_1) + b_2 \cdot (As_2 + e'_2) + e_u (\mathrm{mod} q) \, .
\end{aligned} \quad (9)$$

Here $f$ is still a 1-to-1 function, however $g$ is 4-to-1 so that the entropy difference for these circuits will be 2. This construction can be generalized so that, for any constant $k$, the function $g^{(k)}$ can be made into a $2^k$-to-1 function which corresponds to the entropy difference problem with separation $k$. Hence, we can map the entropy difference problem with arbitrary separation into a similar problem of LWE.

To make connections to a more physical setup, we define the Hamiltonian Quantum Entropy difference problem, following [1].

**Definition 4** *(HQED) Let $H_1$ and $H_2$ be local Hamiltonians acting on $N + K$ qubits, whose ground states are $|\psi_1\rangle$ and $|\psi_2\rangle$. Define the following $N$-qubit mixed states:*

$$\rho_1 = \mathrm{Tr}_K(|\psi_1\rangle\langle\psi_1|) \, , \quad \rho_2 = \mathrm{Tr}_K(|\psi_2\rangle\langle\psi_2|) \, . \quad (10)$$

*Given $N$, $K$ and descriptions of $H_1$ and $H_2$ as input, decide whether:*

$$S(\rho_1) \geq S(\rho_2) + 1 \, , \quad (11)$$

*or*

$$S(\rho_2) \geq S(\rho_1) + 1 \, , \quad (12)$$

*promised that one of these is the case. For the cases where either of the two Hamiltonians has a degenerate ground space, $|\psi_j\rangle$ will denote a state in the ground space for which $S(\rho_j)$ is minimal.*

The theorem proved in [1] states that this Hamiltonian quantum entropy difference problem is in the same complexity class of problem with LWE and the entropy difference problem.

**Theorem 5** *There exists a deterministic poly-time reduction from QED to HQED (QED $\leq_P$ HQED).*

The prove starts by constructing Hamiltonians $H_1$ and $H_2$ such that the respective ground states $|\psi_1\rangle$ and $|\psi_2\rangle$ are Feynman-Kitaev history states for quantum circuits $C_1$ and $C_2$ respectively,

$$|\psi_j\rangle = \frac{1}{\sqrt{T+1}} \sum_{t=0}^{T} U_t U_{t-1} \ldots U_1 |00\ldots 0\rangle |t\rangle \, . \quad (13)$$

where $C_j = U_T U_{T-1} \ldots U_1$, $j \in \{1, 2\}$ and the states $|j\rangle$ are the clock states in the Feynman-Kitaev history state construction. A trick is used to help with the reduction. Given the input circuit $C_j$ to the problem QED, first apply $n$ identity operators at the end of the circuit to each qubit; we will call this the padded circuit. Clearly this does not affect the final state, but it will be useful in the reduction to HQED. We now construct the Feynman-Kitaev history state from the padded circuit, which is

$$\begin{aligned}
|\psi_j\rangle &= \frac{1}{\sqrt{T+n+1}} \sum_{t=0}^{T+n} U_t U_{t-1} \ldots U_1 |00\ldots 0\rangle |t\rangle \\
&= \frac{1}{\sqrt{T+n+1}} \left( U_T U_{T-1} \ldots U_1 |00\ldots 0\rangle \sum_{t'=T}^{T+n} |t'\rangle + \sum_{t=0}^{T-1} U_t U_{t-1} \ldots U_1 |00\ldots 0\rangle |t\rangle \right) \, ,
\end{aligned} \quad (14)$$

because all unitaries $U_i$ for $T+1 \leq i \leq T+n$, $U_i = I^{\otimes(N+K)}$. By the Feynman-Kitaev construction there is a local Hamiltonian, having $poly(N + K)$ terms, for which this is a ground state. Denote $|\phi_j(t)\rangle = U_t U_{t-1} \ldots U_1 |00\ldots 0\rangle$ (with $|\phi_j(0)\rangle = |00\ldots 0\rangle$), then

$$|\psi_j\rangle = \frac{1}{\sqrt{T+n+1}} \left( |\phi_j(T)\rangle \sum_{t'=T}^{T+n} |t'\rangle + \sum_{t=0}^{T-1} |\phi_j(t)\rangle |t\rangle \right) . \tag{15}$$

Note that $|\phi_j(T)\rangle$ is the output state of the circuit $C_j$. If we now trace out the clock register, we have

$$\sigma_j = \mathrm{Tr}_t \left( |\psi_j\rangle \langle \psi_j| \right) = \frac{n+1}{T+n+1} |\phi_j(T)\rangle \langle \phi_j(T)| + \frac{1}{T+n+1} \sum_{t=0}^{T-1} |\phi_j(t)\rangle \langle \phi_j(t)| . \tag{16}$$

One can then calculate the fidelity:

$$F(\sigma_j, |\phi_j(T)\rangle \langle \phi_j(T)|) = \langle \phi_j(T)| \sigma_j |\phi_j(T)\rangle = \frac{n+1}{T+n+1} + \frac{1}{T+n+1} \sum_{t=0}^{T-1} |\langle \phi_j(t) | \phi_j(T)\rangle|^2$$
$$\implies F(\sigma_j, |\phi_j(T)\rangle \langle \phi_j(T)|) \geq \frac{n+1}{T+n+1} = 1 - \frac{T}{T+n+1} . \tag{17}$$

One can then explore the relation between the fidelity and the trace distance to obtain a bound on the trace distance after tracing out the $k$ qubits from both of these states and use the fact that this is a trace non-increasing operation

$$TD\left(\rho'_j, \rho_j\right) \leq \frac{1}{\mathrm{poly}(N+K)} , \tag{18}$$

where $\rho_j$ is the output state of $C_1$ when tracing out the subsystem of $k$ qubits and $\rho'_j$ is the analogous state for the Hamiltonian $H_j$. Then apply the Fannes-Audanaert inequality [4, 5] relating trace distance and entropy. We then would have:

$$\left| S\left(\rho'_j\right) - S\left(\rho_j\right) \right| \leq \frac{1}{\mathrm{poly}(N+K)} . \tag{19}$$

By the triangle inequality we get that if $S(\rho_1) \geq S(\rho_2) + \mathcal{O}(1)$, then $S(\rho'_1) \geq S(\rho'_2) + \mathcal{O}(1)$ and if $S(\rho_2) \geq S(\rho_1) + \mathcal{O}(1)$, then $S(\rho'_2) \geq S(\rho'_1) + \mathcal{O}(1)$. Thus, from the instance of QED $(N+K, K, C_1, C_2)$ we have constructed an instance of HQED $(N+K+n+T, K+n+T, H_1, H_2)$ that preserves the entropy difference of the original circuits (up to a $1/\mathrm{poly}(N+K)$ error).

### A. Holographic attacks against LWE: reading the fine print

We now attempt to analyze this problem through the lens of holographic duality [6]. A promising route appears to emerge from the Ryu-Takayanagi (RT) formula [7], which states that, **to leading order**, the entanglement entropy of a boundary CFT region is given by the area of a minimal surface $\gamma_A$ in the bulk that is homologous to the region:

$$S_A = \frac{\mathrm{Area}(\gamma_A)}{4G_N} . \tag{20}$$

Due to the fact that the AdS/CFT correspondence is a strong-weak duality, in a way, this duality allows us to solve difficult problems in a strongly coupled CFT efficiently in the bulk where everything is weakly coupled. The Ryu-Takayanagi formula is an example of this, which relates a quantity that is hard to compute—entanglement entropy, known to be intractable even for shallow quantum circuits—to a seemingly tractable geometric quantity: the area of a surface. Thus, assuming the hardness of Learning With Errors (LWE) for polynomial-time quantum computers, this correspondence appears to present a contradiction.

A related thought experiment was proposed in [1]. Suppose that the CFT states can be prepared from classical descriptions of functions $f$ and $g$ (for instance, ETFC functions as defined in Eqn. (8)). As previously discussed, we can define these functions such that the resulting states exhibit an arbitrarily large difference in entanglement entropy. It is important to emphasize that the Ryu–Takayanagi (RT) formula captures only the leading-order term in the entropy, which scales as $\mathcal{O}(N)$, where $N$ denotes the number of degrees of freedom of the CFT—in other words, the number of qubits in its discretized form.

Accordingly, the following thought experiment will **not** resolve the HQED problem in the case where the two states differ in entropy by only $\mathcal{O}(1)$. In such cases, one must consider subleading corrections to the RT formula, as discussed in [8]. Both scenarios—entropy differences of $\mathcal{O}(N)$ and $\mathcal{O}(1)$—are physically and computationally significant for the CFT side as stated in Thm. 5, and we will analyze them separately in the sections to follow.

For the purposes of this section, we assume that $f$ is one-to-one and $g$ is $2^{\alpha N}$-to-one, where $\alpha \in (0,1)$ is a fixed constant, such that $\alpha N = k \in \mathbb{Z}^+ \sim \mathcal{O}(N)$.

The thought experiment proceeds as follows. Suppose Alice is given a function $h$, promised to be either $f$ or $g$, and is tasked with determining which type $h$ is. Due to injective invariance, distinguishing between these two cases is as hard as solving LWE. However, Alice may attempt to leverage holography:

- She first prepares a boundary state $|\psi_h\rangle_{\text{CFT}}$, which is assumed to be efficiently generable from the description of $h$. If $h = f$, the state has high entanglement entropy; if $h = g$, it has low entropy.

- Via the AdS/CFT correspondence, $|\psi_h\rangle_{\text{CFT}}$ is dual to a bulk state $|\psi_h\rangle_{\text{bulk}}$. The RT formula then implies that the area of the corresponding minimal surface $\gamma_h$ in the bulk equals the entanglement entropy of $|\psi_h\rangle_{\text{CFT}}$. Alice imagines a bulk observer evolving under a Hamiltonian $H_{\text{bulk}}$, designed such that the observer measures the area of $\gamma_h$ and reacts accordingly: if the area is above a threshold (indicating high entropy), the observer resets itself, returning the system to $|\psi_h\rangle_{\text{bulk}}$. If the area is below the threshold, the observer evolves the system into a bulk state dual to one that is as distinguishable as possible from $|\psi_h\rangle_{\text{CFT}}$.

- Using the AdS/CFT dictionary, Alice derives the boundary Hamiltonian $H_{\text{CFT}}$ dual to $H_{\text{bulk}}$ and evolves her state $|\psi_h\rangle_{\text{CFT}}$ accordingly. This process simulates the bulk observer's dynamics on her quantum computer.

- After evolution, Alice performs a SWAP test to determine whether her final state is still $|\psi_h\rangle_{\text{CFT}}$. If so, she concludes $h$ was 1-to-1; otherwise, she concludes $h$ was 2-to-1.

If all of the above steps could be carried out efficiently, Alice would possess a quantum algorithm that violates the injective invariance of ETCF functions, thereby yielding an efficient solution to LWE. This would contradict standard cryptographic assumptions and challenge the quantum-extended Church–Turing thesis (QECTT). On the gravitational (AdS) side of the holographic duality, this appears unproblematic: the entanglement entropy of a boundary region can be computed geometrically using the Ryu-Takayanagi (RT) formula by evaluating the area of a minimal surface in the bulk. However, from the CFT perspective, estimating the entropy of a general quantum state is computationally hard.

There are three proposed possibilities to explain this:

- **Holographic duality is computationally hard.** Directly accessing bulk geometry requires solving the dual boundary quantum many-body problem, maintaining the problem's intrinsic hardness. Some evidence along this path is shown in [9, 10].

- **Learning with Errors is easy.** If LWE were tractable, the entropy difference problem would become easy. But this contradicts widely held beliefs in cryptography and is considered implausible.

- **Violation of the quantum-extended Church-Turing thesis.** If bulk area measurements are physically efficient but quantumly intractable to simulate, this would challenge the thesis that all physical processes can be efficiently modeled by quantum computers [11].

In this work, we highlight a **fourth possibility**. Precise measurements of areas in AdS space are more subtle than they may appear, since the AdS curvature radius is a global parameter whose influence pervades all large-scale length measurements in the vacuum geometry. For a genuinely local length measurement, the required precision scales only with the size of the segment under consideration, as locally AdS space is indistinguishable from flat space, as any smooth manifold must be. However, the situation is markedly different when computing entanglement entropy, where the relevant minimal geodesics are inherently global objects. Their lengths depend sensitively on the curvature radius $R$. We show that obtaining this information with high precision requires an exponentially large number of measurements. Consequently, the task remains computationally hard even from the bulk perspective, and thus there is no need to assume intrinsic hardness of the AdS/CFT correspondence in order to reconcile with the Quantum Extended Church–Turing Thesis (QECTT).

To make this argument more concrete, consider the problem of estimating entanglement entropy differences between two CFT states. In the $\text{AdS}_3/\text{CFT}_2$ correspondence, the leading-order entanglement entropy is given by the Ryu-Takayanagi formula. However, this is the same as the universal term for the vacuum 1+1d CFT entropy expressed purely in terms of the central charge,

$$S_A = \frac{c}{3} \log\left(\frac{\ell}{\epsilon}\right) + \cdots , \tag{21}$$

where $\ell$ is defined by the subregion one choose. This is derived soly from the understanding of the symmetry of 2 dimensional CFT [12]. We demonstrate that these are the same by analytically deriving the Brown-Henneaux central charge $c$ [13] from the geodesic equation in AdS$_3$.

Consider pure AdS spacetime as a reference background. Given that the entanglement entropy difference between the two candidate boundary states is small—on the order of a single bit—it is justifiable to neglect subleading corrections to the bulk geometry. Let us denote the boundary CFT state as $|\psi_h\rangle$, defined on a subregion $A \subset \partial \text{AdS}$. This subregion is entangled with its complement $\bar{A}$, and the entanglement entropy $S_h$ is holographically captured by the area of the corresponding Ryu-Takayanagi surface in the bulk.

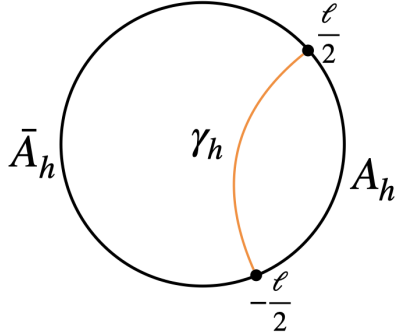

Figure 2. Boundary part $A_h$ corresponding to the CFT state $|\psi_h\rangle$ and its complementary part $\bar{A}_h$.

The Poincaré patch of AdS$_3$ has the metric:

$$ds^2 = \frac{R^2}{z^2}\left(-dt^2 + dx^2 + dz^2\right) , \tag{22}$$

where $R$ is the AdS radius, $z$ is the bulk radial coordinate ($z=0$ is the AdS boundary), and $x$ is the spatial coordinate of the CFT on the boundary. A static extremal surface that computes the entanglement entropy of an interval on the boundary at $t=0$ is given by a *spacelike* geodesic in the bulk.

Consider a boundary interval $A_h = [-\ell/2, \ell/2]$ in the CFT at $z=0$. The corresponding RT surface is a minimal geodesic in the bulk connecting the endpoints of this interval. Since the AdS metric does not explicitly depend on $x$, the geodesic equation:

$$\frac{d^2 x^\mu}{d\tau^2} + \Gamma^\mu_{\rho\sigma} \frac{dx^\rho}{d\tau} \frac{dx^\sigma}{d\tau} = 0 , \tag{23}$$

tells us that there is an invariant quantity:

$$\frac{dx}{d\tau} = C \cdot \frac{z^2}{R^2} . \tag{24}$$

More explicitly, the "Lagrangian" that is used to determine the geodesic equation is not explicitly dependent on $x$. Hence, since $x$ is cyclic, the conjugate momentum will be a conserved quantity which is presented above. Then, We can reparametrize the geodesic so that $x$ is an explicit function of $z$:

$$x = x(z) . \tag{25}$$

The induced metric on the AdS time slice is:

$$ds^2_{\text{induced}} = \frac{R^2}{z^2}\left(dx^2 + dz^2\right) . \tag{26}$$

We then obtained the following equation:

$$\left(\frac{dx}{d\tau}\right)^2 + \left(\frac{dz}{d\tau}\right)^2 = \frac{E^2 z^2}{R^2} , \tag{27}$$

where $E$ is another constant along the geodesic, which is related to the total energy of the geodesic. Eqn. (24) and Eqn. (27) gives the following equation:

$$\left(\frac{dx}{dz}\right)^2 = \frac{C^2 z^2}{R^2 \left(E^2 - C^2 z^2 / R^2\right)} . \tag{28}$$

The geodesic has a turning point $z_*$ where $dz/d\lambda = 0$, which gives:

$$\frac{E^2 z^2}{R^2} - \left(\frac{C z^2}{R^2}\right)^2 = 0 \Rightarrow E^2 = \frac{C^2 z_*^2}{R^2} \Rightarrow C = \frac{ER}{z_*}. \tag{29}$$

This $z_*$ is **the maximum depth of the geodesics.** We can use $z_*$ to express $dx/dz$:

$$\frac{dx}{dz} = \frac{z}{\sqrt{z_*^2 - z^2}}. \tag{30}$$

From which we integrate to get the full geodesic function:

$$x = \pm z_* \sqrt{1 - \frac{z^2}{z_*^2}}. \tag{31}$$

Now we apply the boundary conditions. The boundary interval on the CFT is $A_h = [-\ell/2, \ell/2]$. We also need to apply some UV cutoff: $\epsilon$ for the $z$ coordinate. Applying the boundary condition and solve for the turning point $z_*$:

$$x(z = \epsilon) = \pm \frac{\ell}{2} \quad \Rightarrow \quad \lim_{\epsilon \to 0} z_* \sqrt{1 - \frac{\epsilon^2}{z_*^2}} = \frac{\ell}{2} \quad \Rightarrow \quad z_* \approx \frac{\ell}{2}. \tag{32}$$

Hence, we have the geodesic function $x(z)$:

$$x(z) = \pm \frac{\ell}{2} \sqrt{1 - \frac{4z^2}{\ell^2}}. \tag{33}$$

For small $z$, we have:

$$x(z) \approx \pm \frac{\ell}{2}\left(1 - \frac{2z^2}{\ell^2}\right) = \pm \frac{\ell}{2} - \frac{z^2}{\ell}. \tag{34}$$

We can rewrite the proper length integral as the following:

$$L = 2R \int_\epsilon^{z_*} \frac{dz}{z\sqrt{1 - (z/z_*)^2}} = 2R \log \frac{\sqrt{z_*^2 - \epsilon^2} + z_*}{\epsilon} \approx 2R \log \frac{2z_*}{\epsilon} = 2R \log \frac{\ell}{\epsilon}, \tag{35}$$

again, $\epsilon$ is the UV cutoff of $z$ and $z_* = \ell/2$ is the turning point. We now apply the RT formula [7]. The entanglement entropy is given by:

$$S_h = \frac{L}{4G_N} = \frac{c}{3} \log \frac{\ell}{\epsilon}, \tag{36}$$

where $c$ is the central charge defined before. Hence, given that the central charge is holographically related to the AdS curvature radius via

$$c = \frac{3R}{2G_N}, \tag{37}$$

knowledge of the full bulk geometry implies knowledge of the curvature radius $R$, and hence of $c$.

The problem here is then since the CFT is strongly coupled, the central charge is not simply obtained from the CFT side. However, a natural question is whether this problem can be addressed within the framework of AdS/CFT. At this stage, we know that the system is in the ground state of a CFT, and by consulting the dictionary we may infer that the dual bulk geometry corresponds to pure AdS space. What remains undetermined, however, is the curvature radius of AdS. The central charge of the boundary CFT is directly related to this radius, but in a strongly coupled regime the central charge is generally inaccessible from the Hamiltonian alone. If one accepts the principle underlying the Quantum Extended Church–Turing Thesis (QECTT), it is unreasonable to expect that the AdS curvature radius could be extracted simply by consulting the dictionary. The question then arises: can this radius be determined efficiently through bulk measurements or reconstruction procedures? In this work, we argue that the answer is negative.

In effect, we can view the holographic correspondence as an operational protocol that mediates between two observers: one residing in the CFT and the other in the emergent bulk. When a boundary observer efficiently prepares a CFT state (e.g., via a quantum circuit), a dual bulk geometry is induced. The corresponding bulk observer, located at the conformal boundary, then carries out physical measurements—such as estimating the RT surface length—to extract meaningful geometric information. The mapping between the two observers can thus be interpreted as a communication channel: efficient state preparation on the boundary corresponds to preparing a bulk configuration that encodes the same information, albeit in a geometrized language. In this framework, we will then be able to show that measuring the leading-order entropy is computationally hard due to the exponential number of bulk measurements required. This gives an alternative perspective since the difficulty is not circumvented by formal knowledge of the dictionary, but rather arises from the physical act of measurement within the bulk spacetime.

As we will demonstrate in subsequent sections, estimating the geodesic length corresponding to the RT surface—necessary even for the leading-order entropy—requires resolving distances at the Planck scale. In particular, when using physical probes such as heavy fields, one must perform exponentially many measurements in $N$, the number of degrees of freedom in the boundary theory, to achieve the required precision. Thus, despite being semiclassical, the RT term does not correspond to a tractable observable in any generic, physically realizable bulk protocol.

Beyond the RT contribution, subleading $\mathcal{O}(1)$ corrections arise from bulk quantum entanglement across the RT surface and gravitational loop effects [8]. These quantum corrections require computing the von Neumann entropy of the reduced bulk state in the entanglement wedge—a task that, in the absence of special symmetries or simplifications, also demands exponential computational resources in $N$.

On the CFT side, these difficulties are mirrored. As shown in [1] using cryptographic arguments based on the Learning with Errors (LWE) problem, estimating entropy beyond the universal leading term is computationally intractable for a general class of states. Subleading corrections depend on non-universal data such as the full operator spectrum, OPE coefficients, and conformal block structures. Accessing this information in strongly coupled CFTs involves manipulating correlation functions in Hilbert spaces of size $2^N$, reinforcing the exponential cost of fine-grained entropy estimation.

Crucially, our proposal builds on and extends these prior insights. We emphasize that the difficulty lies not merely in accessing quantum corrections or state-dependent data, but in the very act of measuring the classical RT surface itself—when interpreted as a physical observable in the bulk. To sharpen this point, we focus on observers located at the conformal boundary of AdS spacetime. This assumption is not merely a simplification: in curved geometries, measurable quantities are fundamentally observer-dependent, and defining local bulk observables requires anchoring them to the asymptotic boundary [14, 15].

Taken together, these results support a unified and strengthened conclusion: ***Estimating entanglement entropy in holographic settings is generically computationally hard—even at leading order.*** The apparent simplicity of the RT formula conceals the physical and computational complexity of extracting its inputs, whether from the bulk or boundary perspective. Rather than providing an efficient decoding map for strongly coupled systems, holography should be understood as a correspondence between observers in dual theories, constrained by shared limits on information access and computational power.

## II. OPERATIONAL ESTIMATION OF HOLOGRAPHIC ENTROPY AT LEADING ORDER

To analyze the operational difficulty of estimating entanglement entropy in AdS/CFT, we begin with the semiclassical Ryu–Takayanagi (RT) prescription in AdS$_3$/CFT$_2$. This case offers a tractable setting where both the bulk and boundary quantities are analytically controlled. The central charge $c$ of the boundary CFT, a measure of its degrees of freedom, is related to the bulk Newton's constant $G_N$ and the AdS curvature radius $R$ via the Brown–Henneaux formula [13]:

$$c = \frac{3R}{2G_N} \sim N \,, \tag{38}$$

where $N$ denotes the number of boundary degrees of freedom, or equivalently, the number of qubits in a discretized CFT.

From Eqn. (38), the bulk Newton's constant scales inversely with $N$,

$$G_N \sim \frac{1}{N} \,, \tag{39}$$

and the Planck length in 2+1 dimensions similarly satisfies:

$$\ell_{\mathrm{Pl}} \sim G_N \sim \mathcal{O}\left(\frac{1}{N}\right) \,. \tag{40}$$

Applying the RT formula to compute the entanglement entropy of an interval of length $\ell$ on the boundary yields:

$$S = \frac{L_{\rm RT}}{4G_N} \sim \mathcal{O}(N) \quad \Rightarrow \quad L_{\rm RT} \sim \mathcal{O}(1) \;, \tag{41}$$

since the entropy of a single interval scales linearly with $c \sim N$, and hence the geodesic length $L_{\rm RT}$ must be of order unity.

To estimate the entanglement entropy up to leading order in the bulk, one needs to measure the length of the corresponding RT geodesic with $\mathcal{O}(1)$ precision. This process lies squarely within the semiclassical regime and, in principle, can be performed using classical geometric probes. In contrast, resolving subleading corrections to the entropy—on the order of a single bit—requires significantly greater precision. Specifically, as discussed above, one must resolve the geodesic length to within $\delta L_{\rm RT} \sim \mathcal{O}(1/N)$, which is parametrically equal to the Planck length in three bulk dimensions. This transition in required measurement precision demarcates the onset of quantum gravitational effects. Since at the Planck scale, the quantum structure of spacetime can no longer be neglected. Importantly, this quantum sensitivity is not merely a property of the probing apparatus, but of the bulk geometry itself.

To wit, consider the analytic expression for the geodesic length in AdS$_3$ from Eqn. (35):

$$L_{\rm RT} = 2R \log\left(\frac{\ell}{\epsilon}\right) \;. \tag{42}$$

Here, the boundary interval length $\ell$ and UV cutoff $\epsilon$ are fixed by boundary conditions: they depend only on the spatial subregion under consideration in the boundary theory and on the regulator used in defining the CFT state. Consequently, any variation in $L_{\rm RT}$ that is not due to boundary choices must arise from variation in the AdS curvature radius $R$.

Therefore, resolving differences in geodesic length to $\mathcal{O}(1/N)$ accuracy effectively entails distinguishing between AdS geometries with slightly different curvature radii. In this sense, the emergence of Planck-scale sensitivity is not merely a feature of the local measurement protocol, but rather a manifestation of the quantum nature of the AdS spacetime itself. The inaccessibility of subleading entropy corrections thus encodes genuinely quantum gravitational effects—rooted not in perturbative field excitations, but in the quantization of the background geometry. These effects will be analyzed in more detail in later sections.

### A. Geodesic length from bulk two point functions

The standard method for estimating geodesic lengths in Anti-de Sitter (AdS) spacetime is via two-point correlation functions of bulk fields. To build intuition, we review the semiclassical argument in Euclidean signature, where the two-point function of a heavy scalar field on a fixed background is dominated, in the large-mass limit, by the exponential of minus the geodesic length between the insertion points. Let $\phi(x)$ denote a real scalar field of mass $m$ propagating on a Euclidean manifold with metric $g_{\mu\nu}$, governed by

$$S[\phi] = \frac{1}{2}\int d^dx \sqrt{g}\left(g^{\mu\nu}\partial_\mu\phi\partial_\nu\phi + m^2\phi^2\right) \;. \tag{43}$$

The Euclidean two-point function takes the standard path integral form

$$G(x,y) = \langle\phi(x)\phi(y)\rangle = \frac{1}{Z}\int \mathcal{D}\phi\; \phi(x)\phi(y)e^{-S[\phi]} \;, \tag{44}$$

which, for a free theory, is equivalent to the Green's function of the operator $-\nabla^2 + m^2$, satisfying

$$\left(-\nabla^2 + m^2\right)G(x,y) = \frac{1}{\sqrt{g}}\delta^{(d)}(x-y) \;. \tag{45}$$

Instead of solving this PDE directly, one can employ the Schwinger proper-time representation [16] to express the Green's function as a path integral over relativistic particle trajectories. Formally,

$$G(x,y) = \langle x|\frac{1}{-\nabla^2 + m^2}|y\rangle = \int_0^\infty ds \langle x|e^{-s\left(-\nabla^2+m^2\right)}|y\rangle \;. \tag{46}$$

This admits a path-integral representation as a quantum-mechanical transition amplitude from $y$ to $x$ in Euclidean time $s$ [17, 18]:

$$G(x,y) = \int_0^\infty ds \int_{x(0)=y}^{x(s)=x} \mathcal{D}x^\mu(\tau) \exp\left[-\int_0^s d\tau \, \frac{1}{2}\left(\dot{x}^2 + m^2\right)\right] . \tag{47}$$

In the large-$m$ limit, the integrand is sharply peaked around the classical trajectory that minimizes the action-the geodesic $\gamma_{xy}$ connecting $x$ and $y$ [19, 20]. A saddle-point approximation then yields the dominant contribution

$$G(x,y) \sim e^{-mL_{\text{geo}}(x,y)} , \tag{48}$$

where $L_{\text{geo}}(x,y)$ is the proper length of the geodesic. This geodesic approximation is valid provided $m \gg 1/L_{\text{geo}}$, the field is weakly coupled, and the geodesic is unique and non-degenerate.

To go beyond leading order, we invoke the WKB (Wentzel-Kramers-Brillouin) approximation, which provides an asymptotic expansion in $1/m$ [21, 22]. We seek a solution of the homogeneous equation

$$\left(-\nabla^2 + m^2\right) G(x,y) = 0 \quad (x \neq y) , \tag{49}$$

of the form

$$G(x,y) = e^{-m\sigma(x,y)} \sum_{n=0}^\infty \frac{A_n(x,y)}{m^n} , \tag{50}$$

where $\sigma(x,y)$ is a scalar function interpretable as an effective geodesic distance, and $A_n(x,y)$ are smooth coefficient functions. Substituting into the differential equation and applying the Laplacian term-by-term yields

$$\nabla^2 G = e^{-m\sigma} \sum_{n=0}^\infty \frac{1}{m^n} \left[-m\nabla^2\sigma \cdot A_n + m^2(\nabla\sigma)^2 A_n - 2m\nabla^\mu\sigma\nabla_\mu A_n + \nabla^2 A_n\right] . \tag{51}$$

Equating powers of $m$, we obtain the hierarchy $\left(-\nabla^2 + m^2\right) G = e^{-m\sigma} \sum_{n=0}^\infty \frac{1}{m^n} \mathcal{E}_n = 0$, where

$$\begin{aligned}
\mathcal{E}_0 &= \left[(\nabla\sigma)^2 - 1\right] A_0 \\
\mathcal{E}_1 &= \left[-\nabla^2\sigma \cdot A_0 - 2\nabla^\mu\sigma\nabla_\mu A_0\right] + \left[(\nabla\sigma)^2 - 1\right] A_1 \\
\mathcal{E}_2 &= \left[-\nabla^2\sigma \cdot A_1 - 2\nabla^\mu\sigma\nabla_\mu A_1 + \nabla^2 A_0\right] + \left[(\nabla\sigma)^2 - 1\right] A_2 \\
&\vdots
\end{aligned} \tag{52}$$

and so forth. Setting each $\mathcal{E}_n = 0$ in turn determines the expansion coefficients.

The zeroth-order condition yields the Hamilton-Jacobi equation

$$(\nabla\sigma)^2 = 1 , \tag{53}$$

identifying $\sigma(x,y) = L_{\text{geo}}(x,y)$ as the geodesic distance. This is becauyse the action of a free massive particle traveling along a path $x^\mu(\lambda)$ from $y$ to $x$ is

$$S[x(\lambda)] = m \int_0^1 d\lambda \sqrt{g_{\mu\nu} \frac{dx^\mu}{d\lambda} \frac{dx^\nu}{d\lambda}} = mL_{\text{geo}}(x,y) , \tag{54}$$

and is extremized when the path is a geodesic. The Hamilton-Jacobi equation is the following:

$$g^{\mu\nu}\partial_\mu S \partial_\nu S = m^2 , \tag{55}$$

which we can easily match this with our $\sigma$:

$$S = mL_{\text{geo}} = m\sigma . \tag{56}$$

Then the on-shell action then directly reproduces the classical contribution.

To determine the prefactor $A_0(x,y)$, we consider the first-order transport equation arising from $\mathcal{E}_1 = 0$:

$$2\nabla^\mu\sigma\nabla_\mu A_0 + \left(\nabla^2\sigma\right) A_0 = 0 \tag{57}$$

which has solution

$$A_0(x,y) = [\Delta(x,y)]^{1/2} \tag{58}$$

where $\Delta(x,y)$ is the Van Vleck-Morette determinant, encoding the fluctuation determinant of nearby geodesics. Higher-order terms $A_n$ can be computed recursively via the relation

$$2\nabla^\mu \sigma \nabla_\mu A_n + (\nabla^2 \sigma) A_n = \nabla^2 A_{n-1} \ . \tag{59}$$

This expansion provides a controlled semiclassical series for the Green's function in powers of $1/m$, with the leading exponential decay directly governed by the geodesic length between the operator insertions. As such, the measurement of two-point correlation functions of massive scalar fields in the bulk serves as a practical and physically meaningful protocol for extracting geometric data, such as RT geodesic lengths, within the semiclassical regime.

### B. Measurement complexity at leading order

As discussed in previous sections, estimating the Ryu–Takayanagi (RT) geodesic length to within $\mathcal{O}(1)$ precision suffices to compute the leading-order entanglement entropy in the AdS bulk. This can be operationally achieved by measuring the bulk two-point function $\langle \phi(X_1)\phi(X_2)\rangle$ of a heavy scalar field in a fixed background geometry. In this appendix, we analyze the fundamental limitations of such measurement procedures using tools from quantum estimation theory, particularly focusing on the scaling of measurement precision with the number of resources [23, 24].

Let $O$ be an observable—such as $\phi(X_1)\phi(X_2)$—measured independently on $M$ identically prepared systems. The sample mean $\bar{O}$ has variance:

$$\text{Var}(\bar{O}) = \frac{1}{M^2} \sum_{i=1}^{M} \text{Var}(O_i) = \frac{\sigma_O^2}{M} \ , \tag{60}$$

assuming the measurements are identically distributed with variance $\sigma_O^2$. The uncertainty in estimating the expectation value $\langle O \rangle$ is then given by

$$\delta O = \sqrt{\text{Var}(\bar{O})} \sim \frac{\sigma_O}{\sqrt{M}} \ . \tag{61}$$

This scaling behavior is known as the *standard quantum limit* (SQL). It characterizes the fundamental statistical noise inherent in quantum measurement, even in semiclassical regimes where backreaction is negligible.

In our setup, the RT geodesic length $L_{\text{geo}}(X_1, X_2)$ is inferred from the bulk correlator of a scalar field with mass $m$, which, in the semiclassical limit, behaves as

$$\langle \phi(X_1)\phi(X_2)\rangle \sim e^{-mL_{\text{geo}}(X_1,X_2)}. \tag{62}$$

Let us now assume we want to resolve an $\mathcal{O}(1)$ difference in geodesic length, $\delta L_{\text{geo}} = (L_{\text{geo}}^f - L_{\text{geo}}^g) \sim \mathcal{O}(1)$. The corresponding variation in the two-point function is then

$$\delta \langle \phi(X_1)\phi(X_2)\rangle \sim e^{-mL_{\text{geo}}^g}(1 - e^{-m\delta L_{\text{geo}}}) \ . \tag{63}$$

Since $L_{\text{geo}}^g$, $L_{\text{geo}}^f$, and $\delta L_{\text{geo}}$ are all of $\mathcal{O}(1)$, the part in the bracket is dominated by 1 if we are using a heavy mass field. Hence we can rewrite this as

$$\delta \langle \phi(X_1)\phi(X_2)\rangle \sim e^{-m \cdot \mathcal{O}(1)} \ . \tag{64}$$

To resolve this signal using independent measurements, the SQL requires that the measurement uncertainty $\delta O \lesssim \delta \langle \phi(X_1)\phi(X_2)\rangle$. Hence, we obtain the condition:

$$\frac{1}{\sqrt{M}} \lesssim e^{-m \cdot \mathcal{O}(1)} \quad \Rightarrow \quad M \gtrsim e^{2m \cdot \mathcal{O}(1)} \ . \tag{65}$$

One important caveat concerns the gravitational backreaction of the probe field on the background geometry. In the regime where the field is sufficiently massive, it effectively behaves as a point particle, and its stress-energy tensor

sources a localized perturbation of the spacetime metric. This perturbation, denoted $h_{\mu\nu}$, satisfies the linearized Einstein equations and scales as

$$h_{\mu\nu} \propto G_N m, \tag{66}$$

where $G_N \sim 1/N$ is the bulk Newton constant in units where the AdS radius is fixed. The resulting correction to the geodesic length is given by

$$\delta L \sim \int_\gamma h_{\mu\nu} u^\mu u^\nu d\lambda \propto G_N m, \tag{67}$$

where the integral is taken along the unperturbed geodesic $\gamma$, and $u^\mu$ is the tangent vector to the geodesic. To ensure that the perturbation $\delta L$ remains negligible and does not interfere with the measurement, we must impose that

$$G_N m \ll 1 \quad \Rightarrow \quad m \ll N. \tag{68}$$

In other words, the probe mass must scale more slowly than linearly with $N$ in the large-$N$ limit. This condition places an upper bound on the mass of the field used in geodesic measurements: while $m \sim \sqrt{N}$ remains viable and ensures small backreaction, choosing $m \sim N$ or larger would introduce an $\mathcal{O}(1)$ change on the length of the geodesic, which will invalidate the approximation we take due to appreciable deformation of the background geometry.

Let us now analyze the implications of the measurement complexity for two different mass regimes.

1. **For $m \sim \sqrt{N}$**: (or similarly $m \sim N^{1/\alpha}$, for $\alpha$ a constant $1 \lesssim \alpha$)

   We assume that we can efficiently prepare this state with $N$ qubit. This is the regime in which the WKB approximation is valid. Assuming $\delta L_{\text{geo}} \sim \mathcal{O}(1)$, we find:

   $$M \gtrsim e^{\mathcal{O}(\sqrt{N})} \quad \text{or} \quad M \gtrsim e^{\mathcal{O}(N^{1/\alpha})}. \tag{69}$$

   Therefore, exponentially many measurements are required to estimate the geodesic length at $\mathcal{O}(1)$ precision. This implies that entropy estimation in the bulk, even at leading order, is computationally inefficient in this regime.

2. **For $m \sim \log N'$**:

   Suppose instead we reduce the mass of the bulk probe such that:

   $$M \gtrsim e^{2m\delta L_{\text{geo}}} \sim e^{\mathcal{O}(\log N')} = \text{poly}(N'). \tag{70}$$

   While this improves the measurement cost, the WKB approximation used to relate the correlator to geodesic length becomes unreliable. To ensure the validity of the WKB expansion, we must maintain $mL_{\text{geo}} \gg 1$. If we take $1/m \sim 10^{-2}$ for robustness, then:

   $$\frac{1}{m} \sim 10^{-2} \quad \Rightarrow \quad m \sim 100. \tag{71}$$

   For $m \sim \log N'$, denote $N$ as the number of qubits used in previous case which is assumed to be efficiently prepared, this implies:

   $$N' \gtrsim e^{100} \sim e^{\sqrt{N}}. \tag{72}$$

   Hence, the number of boundary degrees of freedom must scale as $N' \gtrsim e^{\sqrt{N}}$, implying exponentially many more qubits are needed to construct the boundary state $|\psi_h\rangle$.

   Preparing such a boundary state using an log-depth quantum circuit requires runtime:

   $$T_{\text{prep}} \sim N' \log N' \sim e^{\sqrt{N}} \sqrt{N}, \tag{73}$$

   so if $N'$ itself is exponential in the original input size, the preparation time becomes exponential. Therefore, any gain in measurement efficiency is canceled by exponential overhead in state preparation.

We now summarize the implications of these findings in the form of a theorem.

**Theorem 6** *(Intractability of Entropy Estimation via Ryu-Takayanagi formula at Leading Order)*

*Let $|\psi_h\rangle$ be a holographic CFT state encoding $N$ boundary qubits, dual to a semiclassical AdS geometry. Suppose one seeks to estimate the Ryu–Takayanagi geodesic length $L_{geo}(X_1, X_2)$ to within $\mathcal{O}(1)$ accuracy using two-point functions of a bulk scalar field of mass $m$.*

*Then the following statements hold:*

1. *If $m \sim N^{1/\alpha}$ with $1 \lesssim \alpha$, then the WKB approximation is valid and relates geodesic length to the two-point function. However, achieving $\mathcal{O}(1)$ accuracy requires*

$$M \gtrsim e^{\mathcal{O}(N^{1/\alpha})} \,, \tag{74}$$

   *measurements, which is exponentially costly in the second step of the thought experiment in subsec.I A.*

2. *If $m \sim \log N'$, then $M = \text{poly}(N')$ measurements suffice. However, the WKB approximation remains valid only if*

$$N' \gtrsim e^{\mathcal{O}(\sqrt{N})} \,, \tag{75}$$

   *implying exponentially many qubits are needed in the boundary theory. This undermines the assumption that we can efficiently prepare and evolve this state in the first place. Furthermore, the number of measurements needed again scales exponentially with $N$:*

$$M \gtrsim \text{poly}(e^{\sqrt{N}}) \,, \tag{76}$$

   *where $N$ denotes the number of qubits of a state that can be efficiently prepared.*

All in all, there exists no efficient algorithm in the bulk that can estimate the RT geodesic length up to $\mathcal{O}(1)$ precision in polynomial time. Thus, even the leading-order bulk entropy in holographic duality is generically computationally intractable under reasonable physical assumptions.

### C. Implications on LWE and the quantum-extended Church-Turing thesis

We conclude this section with a discussion of broader implications, particularly for the Learning with Errors (LWE) problem and the quantum-extended Church–Turing thesis (QECTT). These connections further illuminate the fundamental hardness of estimating entropy in quantum gravity, even when holographic duality is available.

To begin, it is important to emphasize that not all quantum states arising in the QED or HQED problems correspond to vacuum states of conformal field theories (CFTs), much less to those with a holographic dual. In complexity theory, a "local" Hamiltonian merely requires that each term acts non-trivially on a finite number of qubits. This definition does not imply spatial locality or enforce any geometric decay of correlations. For a conformal field theory, its vacuum must exhibit specific entanglement structure—most notably, a decay of correlations with distance. The two-point function of local operators with the same conformal dimension satisfies:

$$\langle O_1(x_1) O_2(x_2) \rangle = \frac{C_{12}}{|x_1 - x_2|^{2\Delta}} \,, \tag{77}$$

where $\Delta$ is the conformal dimension. This decaying behavior ensures that distant regions are weakly correlated, which in turn gives rise to the area-law scaling of entanglement entropy at leading order in the bulk. Such spatial structure is absent in general local Hamiltonians relevant for the LWE problem.

Even assuming that one has a CFT ground state with spatially local correlations, it is still nontrivial to determine whether the theory admits a holographic dual. The existence of such a dual depends crucially on the high-energy spectrum of the CFT. In particular, holographic CFTs must have a sparse spectrum at low energies and a dense spectrum at high energies, in order to reproduce the Bekenstein–Hawking entropy of black hole states. This requirement arises because black holes have an entropy proportional to their horizon area, implying an enormous number of microstates in the bulk. A generic CFT does not satisfy this spectral constraint; thus, even among CFTs, holographic duality is highly selective.

This observation has direct consequences for complexity theory. Even if we could design a holographic algorithm that efficiently estimates entropy to $\mathcal{O}(N)$ accuracy. This would imply that we can efficiently distinguish between two states with entropy differing by $\mathcal{O}(N)$ bits, via geodesic-length measurements in the bulk. However, as emphasized above, such an algorithm would only apply to CFTs with a holographic dual and thus would not generalize to

arbitrary local Hamiltonians. The scope of holographic tools is thus fundamentally limited in addressing general quantum complexity-theoretic problems.

Moreover, the preceding subsections detailed a concrete measurement protocol in the bulk: by exploiting the WKB approximation for a heavy field of mass $m \sim \sqrt{N}$, one can extract the geodesic length $L_{\text{geo}}$ between two points $X_1$ and $X_2$ via measurement of the two-point function

$$\langle \phi(X_1)\phi(X_2) \rangle \sim e^{-mL_{\text{geo}}} . \tag{78}$$

By the Ryu–Takayanagi prescription, this geodesic length determines the leading-order entanglement entropy of the dual boundary region. However, as formalized in Theorem 6, there exists **no efficient measurement protocol to estimate $L_{geo}$ to $\mathcal{O}(1)$ precision using such probes.** From the boundary perspective, this translates to the **intractability of estimating leading-order entropy corrections in holographic settings**, which is consistent with the LWE complexity that are widely believed [3, 25].

To build some further rough intuition, consider an oracle separation task involving two Boolean functions $f_1$ and $f_2$ defined on $N$-bit strings. Suppose $f_1$ is one-to-one, and $f_2$ is $2^{\alpha N}$-to-one for some $0 < \alpha < 1$ and $\alpha N \in \mathbb{Z}^+$. This task is equivalent to distinguishing between two quantum states: one with entropy $N$ and the other with entropy $(1-\alpha)N$. Because $f_2$ has exponentially many collisions, distinguishing between these cases requires detecting a collision. The birthday paradox [26, 27] tells us that one must sample $\mathcal{O}(2^{(1-\alpha)N})$ outputs to see such a collision with high probability. Thus, any algorithm solving this entropy-distinguishing task must have exponential runtime in $N$. This is a known hardness reduction in LWE-based cryptography, and the inaccessibility of geodesic-length measurements in the bulk merely reflects this fundamental lower bound. A more rigorous arguments on this can be found in [1], where this fact actually follows from the injective invariance property of the ETCFs.

Finally, we turn to the broader implications for the quantum-extended Church–Turing thesis (QECTT), which asserts that any physically realizable process can be simulated efficiently by a quantum computer. At first glance, one might worry that the semiclassical gravitational process of measuring geodesic lengths in AdS spacetime could violate this thesis. However, our analysis shows the opposite: even in a holographic setup with an efficiently preparable boundary state and an emergent classical bulk, the entropy estimation problem remains computationally intractable. This conclusion rests on the interplay between quantum measurement limits (e.g., the standard quantum limit), backreaction constraints, and the WKB validity regime.

While the AdS geometry is treated semiclassically, the measurement process becomes computationally difficult once quantum corrections become significant. To apply the WKB approximation reliably, one must use probe particles with mass $m \sim \sqrt{N}$. However, pushing the mass beyond this scale induces non-negligible gravitational backreaction, see Eqn. (68), which remains small in the large-$N$ limit but grows with heavier probes, compromising the classical geometry. Alternatively, using lighter probes (e.g., $m \sim \ln N$) reduces measurement cost but breaks the WKB regime. Restoring semiclassical validity in such cases requires exponentially many boundary qubits (e.g., $N' \sim e^{\sqrt{N}}$), making both the state preparation and the measurement intractable. Thus, any attempt to evade exponential cost either invalidates the semiclassical picture or transfers the hardness to state preparation.

In summary, the measurement of leading-order entropy in holographic theories remains computationally hard, both from the boundary and the bulk perspectives. This intractability does not stem from the inaccessibility of the AdS/CFT dictionary itself, but rather from the intrinsic difficulty of the physical operations required to estimate entropy with semiclassical precision. On the boundary, entropy extraction involves nontrivial correlation functions over exponentially large Hilbert spaces. In the bulk, estimating the Ryu–Takayanagi geodesic length to $\mathcal{O}(1)$ accuracy requires exponentially many measurements of heavy field correlators, as shown in Theorem 6, unless one compromises the validity of the semiclassical approximation.

We interpret these findings within a physically grounded formulation of the AdS/CFT correspondence as a duality between operational observers: a boundary observer equipped with quantum circuit tools acting on a CFT Hilbert space, and a bulk observer located at the conformal boundary, performing geometric measurements. This mutual inaccessibility of leading order entropy estimation reinforces the view that holography is not a computational oracle. Rather, it is a dual language for describing physical information, one that is consistent with the quantum-extended Church–Turing thesis (QECTT). In particular, even when a holographic geometry is assumed to be efficiently preparable, the observers within it cannot efficiently perform all computations about it. Holography, then, emerges not as a computational loophole in strongly coupled physics, but as a deep expression of the universality of quantum computational bounds.

## III. MEASUREMENT COMPLEXITY OF QUANTUM CORRECTIONS

In this section we start to study the scenario when we try to estimate the entropy of the ground state of physical local Hamiltonian up to $\mathcal{O}(1)$ accuracy. The motivation for this is two fold. First, the LWE problem, QED problem, and

HQED problem all considered reolving to entropy difference of $\mathcal{O}(1)$, so naturally, we need to include this discussion for completeness. To do this is, it is not enough to simply measuring the length of the geodesic, since as demonstrated at the beginning of sec.II: $\delta L_{\text{geo}} \sim \frac{1}{N} \sim \ell_{\text{Pl}}$. This indicates that precision at this level will require more systematic quantum mechanical treatment, in particular quantum corrections to the RT formula, which, as we will demonstrate in this section, is operational complexity is also exponential in $N$.

Quantum corrections for the RT formula is captured by the Faulkner-Lewkowycz-Maldacena (FLM) prescription [8], which introduce new contributions arising from bulk quantum fluctuations. These corrections are of order $\mathcal{O}(1)$ and, crucially, depend not solely on local geometric quantities but on the entanglement structure of quantum fields across the RT surface. Consequently, measuring these corrections operationally involves accessing highly nonlocal quantum information within the bulk.

In particular, we systematically analyze the measurement complexity associated with quantum gravitational corrections. Our main result can be summarized as follows:

**Theorem 7** *(Measurement complexity of holographic entanglement entropy at $\mathcal{O}(1)$)*

*Extracting the FLM quantum correction to the Ryu-Takayanagi entropy, particularly the bulk entanglement entropy contribution, requires an exponential number of measurements in the number of boundary degrees of freedom $N$. That is, the number of measurements required to reconstruct the bulk entanglement entropy with Planck-scale precision scales as*

$$\mathcal{T}(N) \sim \text{poly}(N) \cdot e^{N^{1/\alpha}} \cdot 2^{\mathcal{O}(N)}, \tag{79}$$

where $1 \lesssim \alpha$.

Importantly, despite this exponential complexity, no violation of the QECTT occurs. Since both the quantum algorithm for solving the entropy difference problem or LWE problem and the holographic algorithm are computationally hard. The exponential difficulty arises for the holographic duality case comes in with reconstructing the full entanglement structure of the bulk fields.

### A. Quantum Corrections to the Ryu-Takayanagi Formula

To start with, we first provide a brief recap of the derivation of the FLM correction terms [8]. The quantum-corrected entanglement entropy of a boundary region $A$ is given by

$$\begin{aligned} S(A) &= S_{\text{cl}}(A) + S_q(A) + \mathcal{O}(G_N), \\ S_q &= S_{\text{bulk-ent}} + S_{...} = S_{\text{bulk-ent}} + \frac{\delta A}{4G_N} + \langle \Delta S_{\text{Wald-like}} \rangle + S_{\text{counterterms}} \end{aligned} \tag{80}$$

where $S_{\text{cl}}(A)$ denotes the classical RT term proportional to the minimal surface area, and $S_{\text{bulk-ent}}(A_b)$ represents the entanglement entropy of bulk quantum fields across the corresponding bulk region $A_b$. The second is the change in the area due to the shift in the classical background due to quantum corrections. The third is the quantum expectation value of the formal expression of the Wald-like entropy. The final term arises because we need to introduce counterterms in order to render the computation finite. A schematic depiction of the bulk division relevant to the quantum correction is shown in Fig. 3.

The replica trick provides a powerful method to compute the von Neumann entropy of a quantum state. For a density matrix $\rho$, the entropy $S$ is given by

$$S = -\text{Tr}(\rho \log \rho) = -\lim_{n \to 1} \partial_n \text{Tr}(\rho^n). \tag{81}$$

In the context of holography, to compute the entanglement entropy of a boundary region $A$, one considers a replicated geometry $g_n$ where $n$ copies of the spacetime are cyclically glued along $A$.

At the quantum level, following the replica trick, the first job is to compute the partition function of all the bulk quantum fields around the black hole geometry. The partition function of bulk quantum fields on this branched cover is denoted $Z_{q,n}$:

$$Z_{q,n} = \text{Tr}\left(\hat{\rho}_n^n\right), \quad \hat{\rho}_n = P \exp\left(-\int_0^{2\pi} H_{b,n}(\tau) d\tau\right), \tag{82}$$

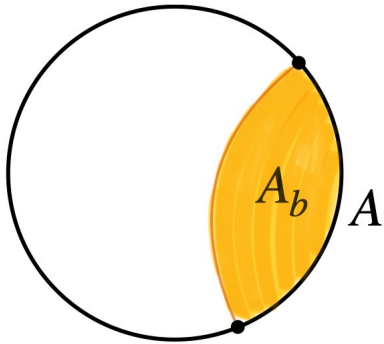

Figure 3. The boundary region $A$ corresponds to a bulk minimal surface dividing the bulk spacetime into two parts. The leading quantum correction arises from the entanglement entropy of bulk quantum fields between the region $A_b$ and its complement.

where $H_{b,n}(\tau)$ is the $\tau$-dependent bulk Hamiltonian, and $P$ denotes path ordering. The quantum correction to the entropy is then extracted via

$$S_q = -\partial_n \left(\log Z_{q,n} - n \log Z_{q,1}\right)\Big|_{n=1}. \tag{83}$$

Expanding around $n = 1$, the result decomposes into the several terms in $S_q$ of Eqn. (80).

### B. Estimating integral of local quantities in the FLM correction

As remarked in ref. [8], the quantum corrections are essentially the bulk entanglement entropy along with some extra one loop terms that can be expressed as an integral of local quantities. In this section, let's first attempt to measure the local quantities involved, which is denoted as $S_{...}$. This $S_{\text{bulk-ent}}$, intuitively, presents a big challenge, as it demands reconstructing fine-grained quantum information within the bulk spacetime, which we leave for next section.

Let's start with the $\Delta S_{\text{Wald-like}}$ term, we consider the structure of gravitational effective field theory. Suppose the low-energy effective action contains generic four-derivative corrections of the form

$$L = c_1 R^2 + c_2 R_{\mu\nu} R^{\mu\nu} + c_3 R_{\mu\rho\nu\sigma} R^{\mu\rho\nu\sigma} . \tag{84}$$

As shown in Ref. [28], the correction to the entanglement entropy due to these higher-derivative terms is given by

$$\delta S_{EE} = -4\pi \int d^d y \sqrt{g} \left[ 2\delta c_1 R + \delta c_2 \left( R_a^a - \frac{1}{2} K_a K^a \right) + 2\delta c_3 \left( R_{ab}^{ab} - K_{aij} K^{aij} \right) \right] , \tag{85}$$

where $K_a$ and $K_{aij}$ are the extrinsic curvatures of the entangling surface.

Since the overall size of $\delta S_{EE}$ is expected to be of order $\mathcal{O}(1)$ in the large-$N$ limit, we infer that

$$\delta c_i \sim \mathcal{O}(R) , \tag{86}$$

where $R$ is the AdS radius. Moreover, as shown in Ref. [29], such higher-derivative corrections can arise from integrating out a heavy scalar field $\phi$ minimally coupled to gravity. The induced couplings are related to the scalar mass by

$$\delta c_i \sim \frac{m_{\text{pl}}}{m_\phi^2} \sim \mathcal{O}(\ell) \quad \Rightarrow \quad m_\phi \sim \sqrt{\frac{m_{\text{pl}}}{\ell}} \sim N^{1/2} . \tag{87}$$

Note that $G_N \sim 1/N$ in $2+1$ spacetime dimensions. Thus, we can integrate out this heavy bulk field coupled to graviton to generating the Wald-like quantum corrections must have a mass parametrically large compared to the AdS curvature scale. At the same time, its mass remain sufficiently small to avoid inducing significant gravitational backreaction on the semiclassical background.

In addition, we can look at the contribution from the counterterms $S_{\text{counterterms}}$, which are local functionals of the geometry (constructed from the metric, curvature, etc.). Thus resemble the terms that already appear in the classical gravitational action. Their effect on the entropy is obtained in the same way as for any higher-derivative correction to the action: they contribute local terms to the entropy functional, evaluated via a Wald-type formula [8]. For this, it is natural to believe that if we can efficiently estimate the Wald-like terms, we can also do the same for the counterterm contributions.

This observation suggests a possible measurement protocol for the Wald-like correction regarding the heavy field $\phi$ as a controllable probe. The key idea is that the one-loop correction to the entanglement entropy is encoded in how virtual $\phi$-loops renormalize local gravitational couplings.

One can imagine measuring the propagation of these $\phi$ fields in AdS - for instance, through boundary two-point or higher-point functions of $\mathcal{O}_\phi$. The deviation of these correlators from pure geodesic WKB scaling can be attributed to curvature-dependent terms, which are precisely the local higher-derivative couplings $\delta c_i$. One can then infer the Wald-like entropy contribution by inserting the extracted $\delta c_i$ into the Wald formula for higher-derivative entanglement entropy, e.g. in Eqn. (85). In this way, the heavy field itself serves as a physical probe of the higher-derivative renormalization of gravity. Similarly, as argued before, the counterterm contributions $S_{\text{counterterms}}$ are likewise local functionals of the geometry, determined by renormalization of the bulk action. Since they enter the entropy via Wald-like formulae [8], the same protocol - using heavy probe fields to detect their effective couplings - can in principle be applied to extract them.

Operationally, both the Wald-like term and the counterterm contributions are inferred from the same set of WKB-corrected boundary correlators. Small changes in the bulk geometry or in local bulk couplings imprint themselves on these correlators through the same smoothing forward map (a geodesic X–ray). In AdS this map suppresses information about deep, transverse modes exponentially, so any protocol that reconstructs the area-sensitive degrees of freedom from boundary two-point data requires exponentially many high-precision measurements as we have demonstrated in sec. II B. **Thus extracting the Wald-like and counterterm pieces from the same correlator data reuses the same data and therefore inherits the same exponential measurement complexity.**

At last we turn to the term $\frac{\delta A}{4G_N}$ in $S_q$. This contribution arises from the change in the area of the RT surface due to the shift in the classical background sourced by quantum fields in the bulk. Since $\delta A \sim G_N$, this term contributes at $\mathcal{O}(1)$.

We work in the semiclassical regime: the bulk contains quantum matter fields with stress tensor $T_{\mu\nu}(y)$ and classical gravity with Newton constant $G_N$. The FLM geometric correction comes from the linearized backreaction of the expectation value $\langle T_{\mu\nu} \rangle$ on the bulk metric. Holding other FLM pieces fixed, we want to measure the geometric entropy contribution

$$\delta S_{\text{geo}} = \frac{\delta \text{ Area}}{4G_N} = \frac{1}{8G_N} \int_A \sqrt{\gamma} \gamma^{ab} h_{ab} d^2 y + \cdots , \tag{88}$$

where $h_{\mu\nu}$ is the metric perturbation linear in $\langle T \rangle$, and $A$ is the classical RT surface (embedding-shift terms are similar). Our operational assumption is that the experimenter is a bulk observer with local access to the quantum fields in a neighborhood containing $A$, and can perform repeated local measurements of matter observables such as the smeared stress tensor.

The linearized Einstein equation takes the schematic form

$$\mathcal{E}[h]_{\mu\nu}(x) = 8\pi G_N \langle T_{\mu\nu}(x) \rangle , \tag{89}$$

where $\mathcal{E}$ is the linearized Einstein operator. Formally inverting with the graviton Green's function $G^{(\text{grav})}$ yields

$$h_{\mu\nu}(x) = 8\pi G_N \int_{\text{bulk}} G^{(\text{grav})\alpha\beta}_{\mu\nu}(x;y) \langle T_{\alpha\beta}(y) \rangle dV_y . \tag{90}$$

Plugging this into the area variation (holding the embedding fixed) gives

$$\delta \text{ Area}(A) = 4\pi G_N \int_{\text{bulk}} \underbrace{\left[ \frac{1}{2} \int_A \sqrt{\gamma} \gamma^{ab} G^{(\text{grav})\alpha\beta}_{ab}(x_A;y) d^2 x_A \right]}_{\equiv \mathcal{K}^{\alpha\beta}(y)} \langle T_{\alpha\beta}(y) \rangle dV_y . \tag{91}$$

Thus the entropy contribution is

$$\delta S_{\text{geo}} = \frac{\delta \text{ Area}}{4G_N} = \pi \int_{\text{bulk}} \mathcal{K}^{\alpha\beta}(y) \langle T_{\alpha\beta}(y) \rangle dV_y , \tag{92}$$

where $\mathcal{K}^{\alpha\beta}(y)$ is a purely geometric kernel of order $\mathcal{O}(1)$ in AdS units. Moreover, this $\delta S_{\text{geo}}$ is of $\mathcal{O}(1)$, so the expectation value $\langle T_{\alpha\beta}(y)\rangle$ is also of $\mathcal{O}(1)$. Crucially, the explicit factor of $G_N$ cancels: $\delta S_{\text{geo}}$ is simply a linear functional of the bulk stress expectation with an $O(1)$ kernel. This structural fact makes the bulk-local measurement operationally tractable.

To design a measurement protocol, discretize the relevant bulk region into $M$ cells labeled by $i$, with volumes $\Delta V_i$. In each cell, measure the (smeared) stress tensor components

$$t_i \equiv \langle T_{\alpha\beta}(y_i)\rangle \ . \tag{93}$$

with the index contraction absorbed into $\mathcal{K}$. The estimator for the entropy is the Riemann sum

$$\widehat{\delta S}_{\text{geo}} = \pi \sum_{i=1}^{M} \mathcal{K}_i \Delta V_i \hat{t}_i \ , \tag{94}$$

where $\mathcal{K}_i \equiv \mathcal{K}^{\alpha\beta}(y_i)$ and $\hat{t}_i$ is the measured sample mean in cell $i$. Assume independent repeated local measurements. Each measurement has single-shot variance $\sigma_i^2$. Averaging $n_i$ independent shots in cell $i$ yields

$$\text{Var}\left(\hat{t}_i\right) = \frac{\sigma_i^2}{n_i} \ . \tag{95}$$

Propagating variance gives

$$\text{Var}\left(\widehat{\delta S}_{\text{geo}}\right) = \pi^2 \sum_{i=1}^{M} (\mathcal{K}_i \Delta V_i)^2 \frac{\sigma_i^2}{n_i} \ . \tag{96}$$

For uniform cells, uniform sampling $n_i = n$, and uniform single-shot variance $\sigma^2$, define

$$\mathcal{I} \equiv \sum_{i=1}^{M} (\mathcal{K}_i \Delta V_i)^2 \approx \int_{\text{bulk}} \mathcal{K}(y)^2 dV_y \ . \tag{97}$$

Then

$$\text{Var}\left(\widehat{\delta S}_{\text{geo}}\right) = \pi^2 \frac{\sigma^2}{n} \mathcal{I} \ . \tag{98}$$

By the standard quantum limit (SQL), to achieve error $\varepsilon$ we require

$$n \gtrsim \frac{\pi^2 \sigma^2 \mathcal{I}}{\varepsilon^2} \ . \tag{99}$$

The total number of shots across all cells is $n_{\text{total}} = Mn$. The cells are concentrated near the classical RT surface, so $M$ scales polynomially with the area (or length in AdS$_3$) of the surface, hence as $\text{poly}(N)$. Likewise, the integral $\mathcal{I}$ is $\mathcal{O}(1)$ in AdS units. Thus the total sample complexity is polynomial in $N$ and in $1/\varepsilon$.

Importantly, there is no exponential suppression in the signal: the bulk-local protocol directly accesses $\langle T_{\mu\nu}\rangle$ without needing to invert the exponentially decaying bulk-to-boundary kernel.

All in all, the measurement complexity of the holographic entanglement entropy at $\mathcal{O}(1)$ beside the $S_{\text{bulk-ent}}$ term can be expressed as a polynomial corrected exponential complexity which was inherited from measuring the RT geodesic length up to $\mathcal{O}(1)$ precision from Thm. 6. Putting these together, we have the following corollary.

**Corollary 8 (Measurement Complexity for estimating $(S_{cl} + S_{...})$ of holographic subregion entropy in AdS$_3$ vacuum)**

*Extracting the classical RT term $S_{\text{cl}}$ together with the Wald-like contributions, the local counterterms, and the $\frac{\delta A}{4G_N}$ term in the holographic entropy at $\mathcal{O}(1)$ requires an exponential number of measurements in the boundary degrees of freedom $N$. This complexity is inherited from Thm. 6, which establishes the exponential cost of measuring the geodesic length $L_{\text{geo}}$ to $\mathcal{O}(1)$ precision, up to additional polynomial overhead from reconstructing local contributions. Concretely, the total measurement cost scales as*

$$\mathcal{T}(N) \sim \text{poly}(N) e^{N^{1/\alpha}} \ , \tag{100}$$

*for some $1 \lesssim \alpha$, with the exponential factor dominating the asymptotics.*

### C. Estimating bulk entanglement entropy via covariance matrices

Now we turn to the estimation of the essential part of bulk quantum correction which is the bulk entanglement entropy between region $A_b$ and the rest of the bulk, denoted as $S_{\text{bulk-ent}}$. In computing this term, we focus on the leading quantum correction, which arises at 1-loop order in bulk effective field theory. At this order, the bulk quantum fields propagate on a fixed classical background geometry (typically pure AdS), and their dynamics are governed by the quadratic part of the effective action. Expanding the bulk action $S[\phi]$ around a classical saddle $\phi = \phi_{\text{cl}} + \delta\phi$ and retaining only the quadratic fluctuations yields

$$S[\phi] \approx S[\phi_{\text{cl}}] + \frac{1}{2}\int d^d x\, \delta\phi(X)\mathcal{K}(X,X')\delta\phi(X') + \cdots, \tag{101}$$

where $\mathcal{K}(X,X')$ is the kernel of the quadratic fluctuation operator. The resulting path integral is Gaussian, and the corresponding quantum state is fully characterized by its two-point function,

$$G(X,X') = \langle \phi(X)\phi(X')\rangle. \tag{102}$$

To make the entanglement structure operationally accessible — both for numerical evaluation and complexity-theoretic analysis — we again work with a discretized version of the bulk theory, in which the spatial geometry of the entanglement wedge is approximated by a lattice of sites. This step mirrors the discretization of the boundary CFT into a system of $N$ qubits or lattice sites, such that the Hilbert space dimension of the boundary subregion $A$ is $\dim \mathcal{H}_A = 2^N$.

In holographic duality, the entanglement wedge $\mathcal{E}_A$ is the bulk region associated with the boundary region $A$, and it is expected that the number of independent bulk degrees of freedom reconstructible from $\mathcal{H}_A$ cannot exceed the number of independent boundary degrees of freedom. More precisely, any bulk operator reconstructible from $A$ must act within the code subspace associated with $\mathcal{H}_A$, which has dimension at most $2^N$. Consequently, in a semiclassical regime where the bulk state is approximated as a Gaussian state, the number of linearly independent, low-energy, canonically normalized field modes that can be reconstructed in $\mathcal{E}_A$ is bounded by $D \sim 2^N$.

We thus discretize $\mathcal{E}_A$ into $D \sim 2^N$ lattice sites, each associated with a pair of canonical variables — the field value $\phi_i$ and its conjugate momentum $\pi_i$. Collectively, we denote these as $\{\xi_i\}$, where each $i$ runs over the $2D$ canonical degrees of freedom. Because the bulk state is Gaussian at 1-loop order, its full entanglement structure is encoded in the covariance matrix

$$\Gamma_{ij} = \langle \xi_i \xi_j + \xi_j \xi_i\rangle - 2\langle \xi_i\rangle\langle \xi_j\rangle, \tag{103}$$

which is a real symmetric $2D \times 2D$ matrix. The number of symplectic eigenvalues $\nu_k$ (i.e., entanglement mode occupation numbers) is $D$, corresponding to the number of independent bosonic modes supported in the bulk entanglement wedge.

The reduced density matrix for a Gaussian state takes the form

$$\rho = \frac{1}{Z}\exp\left(-\frac{1}{2}\sum_{i,j}\xi_i G_{ij}\xi_j\right), \tag{104}$$

where $G$ is a positive-definite matrix related to $\Gamma^{-1}$, and $Z$ is a normalization constant. This reduced state resembles a thermal ensemble of decoupled bosonic modes governed by an effective quadratic Hamiltonian.

The canonical commutation relations impose a symplectic structure, encoded in the matrix $\Omega$, and the fundamental invariants of the Gaussian state are its *symplectic eigenvalues* $\{\nu_i\}$, defined as the spectrum of the matrix $i\Omega^{-1}\Gamma$. These eigenvalues satisfy $\nu_i \geq \frac{1}{2}$ due to the uncertainty principle and encode the entanglement of each mode.

The von Neumann entropy of a single bosonic mode with symplectic eigenvalue $\nu$ is

$$S(\nu) = \left(\nu + \frac{1}{2}\right)\log\left(\nu + \frac{1}{2}\right) - \left(\nu - \frac{1}{2}\right)\log\left(\nu - \frac{1}{2}\right), \tag{105}$$

and the total entanglement entropy is given by

$$S_{\text{bulk-ent}} = \sum_i S(\nu_i). \tag{106}$$

This procedure enables a complete and operationally well-defined computation of the bulk entanglement entropy using only two-point functions, under the assumption that the bulk fields are Gaussian. This approximation is valid

in the 1-loop regime of bulk effective field theory, where quantum corrections arise from free (e.g., scalar or linearized gravitational) fields propagating on a fixed classical background. In this semiclassical limit, interactions and higher-loop effects are suppressed by powers of $1/N$, rendering the bulk state within the entanglement wedge $\mathcal{E}_A$ effectively Gaussian [8, 28].

It is worth noting that in some cases, the leading divergence in the bulk entanglement entropy $S_{\text{bulk-ent}}$ obeys an area law analogous to the Ryu–Takayanagi (RT) formula, it differs in both magnitude and origin. In a semiclassical bulk theory with a UV cutoff $\epsilon$, the entanglement entropy of a vacuum state across the extremal surface $\Sigma$ takes the schematic form:

$$S_{\text{bulk-ent}} = \alpha \frac{\text{Area}(\Sigma)}{\epsilon^{d-2}} + \delta S_{\text{finite}}, \tag{107}$$

where $\alpha$ is a cutoff-dependent constant and $\delta S_{\text{finite}}$ includes all finite contributions (e.g., curvature corrections, long-range entanglement). Unlike the RT term, which scales as $\text{Area}/(4G_N) \sim \mathcal{O}(N)$ in holographic CFTs, the bulk entropy is an $\mathcal{O}(1)$ quantity, as it is independent of $G_N$ at leading order.

In discretized models (e.g., tensor networks or lattice regularizations), the divergent term becomes a large but finite sum over sites near $\Sigma$:

$$S_{\text{bulk-ent}} \sim \sum_{i \in \text{RT layer}} S(\nu_i) + \sum_{j \notin \text{RT layer}} S(\nu_j), \tag{108}$$

where $S(\nu)$ is the entropy per symplectic eigenvalue $\nu$, and the first sum dominates due to area-law scaling. However, we only consider the FLM correction when the entropy *differences* between two semiclassical states cannot be distinguished at $\mathcal{O}(N)$. This implies that the leading term proportional to areas, in our cases, will largely cancel each other:

$$\Delta S_{\text{bulk}} = S_{\text{bulk-ent}}^{(f)} - S_{\text{bulk-ent}}^{(g)} \sim \delta S_{\text{finite}}. \tag{109}$$

Capturing this finite difference $\delta S_{\text{finite}} \sim \mathcal{O}(1)$ requires retaining contributions from many bulk modes, even those with small $\nu_j - 1/2$. These subleading shifts may be individually small but can collectively sum to an appreciable correction. **Therefore, computing $S_{bulk\text{-}ent}$ to order-one accuracy necessitates reconstructing the full set of $\mathcal{O}(2^N)$ bulk modes from the boundary, not just those localized near the RT surface.** Truncating to the dominant area-law terms is insufficient for resolving the FLM correction, even in vacuum AdS.

*1. Measurement Protocol and Complexity Estimate*

In this subsection, we explicitly propose the steps to construct the bulk covariance matrix $\Gamma$ and extract the entanglement entropy, which combined with corollary. 8 will prove the exponential run time stated in Thm. 7.

1. **Construction of Covariance Matrix:** Build $\Gamma_{ij}$ from the measured two-point functions in Eqn. (103).

2. **Symplectic Diagonalization:** Diagonalize $\Gamma$ to obtain symplectic eigenvalues $\{\nu_i\}$.

3. **Entropy Computation:** Compute

$$S_{\text{bulk-ent}} = \sum_i S(\nu_i) . \tag{110}$$

As a result, the measurement complexities for the most general case are summarized below:

| Step | Process | Complexity |
|---|---|---|
| 1 | Build covariance matrix $\Gamma$ | $\mathcal{O}(2^{4N})$ |
| 2 | Diagonalize $\Gamma$ | $\mathcal{O}(2^{6N})$ |
| 3 | Compute entanglement entropy | $\mathcal{O}(2^{2N})$ |

Thus, the bottleneck arises from the exponential size of the full covariance matrix: achieving Planckian precision operationally requires exponential resources, consistent with Theorem 7.

On the other hand, there may be avenues for algorithmic improvement in the estimation of the FLM correction. For free (Gaussian) bulk fields in AdS, the two-point correlators typically decay as a power law with geodesic distance.

This implies that the entries of the bulk covariance matrix $\Gamma$ exhibit polynomial decay away from the diagonal. Explicitly, if $i, j$ index spatial points in the discretized bulk subregion $A_b$, then the correlators scale as [30]:

$$\Gamma_{ij} \sim \frac{1}{|x_i - x_j|^\Delta} , \qquad \alpha > 1 . \tag{111}$$

As a consequence, **one may approximate $\Gamma$ as a sparse or approximately banded matrix, retaining only entries within a geodesic window $|x_i - x_j| \lesssim L$ for some truncation scale $L$.** The resulting approximation $\tilde{\Gamma}$ satisfies

$$|\Gamma - \tilde{\Gamma}| \leq \delta , \tag{112}$$

with $\delta \to 0$ as $L \to \infty$, provided $\Delta > 1$. This allows us to treat the modular operator

$$M := i\Omega^{-1}\Gamma , \tag{113}$$

as a structured matrix amenable to efficient simulation by quantum algorithms. Note that the $\Omega$ is the symplectic form, and we map the symplectic matrix to a Hamiltonian form with its inverse.

In particular, quantum phase estimation (QPE) [31] provides a method to extract the symplectic eigenvalues $\{\nu_k\}$ of $\Gamma$. To estimate these eigenvalues, one encodes $M$ as a Hamiltonian on a quantum register and applies QPE to sample from its spectrum. We assume access to a block-encoding or Trotterized simulation of $e^{-iMt}$, and the ability to prepare quantum states with non-negligible overlap with eigenvectors of $M$. Moreover, this $M$ is $s$ sparse: at most $s = \text{poly}(N)$ nonzero entries per row. Then QPE estimates eigenvalues $\nu_k$ to accuracy $\varepsilon$ with complexity

$$\text{Cost}_{\text{QPE}} = \mathcal{O}\left(\frac{s}{\varepsilon} \text{polylog}(2^{2N})\right) \sim \mathcal{O}\left(\frac{1}{\varepsilon} \text{poly}(N)\right) . \tag{114}$$

The entropy can then be estimated by sampling a sufficient number $J$ of dominant eigenvalues such that

$$\left| S(\rho_{\Sigma_A}) - \sum_{j=1}^{J} s(\nu_j) \right| < \epsilon , \tag{115}$$

where $\epsilon$ bounds the truncation error. The total runtime of the quantum protocol scales as $\mathcal{O}(J/\delta)$, under idealized assumptions.

While this procedure does not necessarily yield a exponential speedup-especially when the symplectic spectrum is broad (when $J \sim 2^N$)-it offers a substantial improvement over brute-force diagonalization, leveraging the approximate locality of correlations in the bulk field theory. In this setting, quantum algorithms for sparse matrix simulation and spectrum estimation (e.g., via blockencoding or linear-combination-of-unitaries techniques) can be deployed [31].

To meaningfully exploit the approximate sparsity of the covariance matrix $\Gamma$, it is necessary to **first compute or estimate the matrix entries, at least in a neighborhood of the diagonal.** This is because the dominant contributions to the entanglement entropy do not simply come from "modes near the RyuTakayanagi (RT) surface" in a geometric sense, but from the symplectic eigenmodes of $\Gamma$ restricted to the entangling region. These eigenmodes are nonlocal linear combinations of the local bulk field modes, and their entropic weight is determined by the full spectral structure of $\Gamma$. Specifically, recall that the entanglement entropy is given by Eqn. (110), then unless one computes the entries of $\Gamma$ — even approximately — one cannot determine *a priori* which linear combinations of modes dominate the entropy.

Moreover, the power-law decay of correlators $\langle \phi(x)\phi(y) \rangle \sim |x - y|^{-\Delta}$ implies that while far-separated points contribute subdominantly, the precise cutoff scale beyond which such contributions can be neglected depends sensitively on the scaling exponent $\alpha$ and the desired accuracy. Truncating too aggressively risks omitting modes with significant weight in the entropy computation. Importantly, the covariance matrix entries encode interference effects between nearby modes, and neglecting these without evaluation may destroy critical correlations.

Therefore, one must first construct or sample a banded approximation $\tilde{\Gamma}$ that includes the neardiagonal structure - i.e., entries $\Gamma_{ij}$ such that the proper distance between $x_i$ and $x_j$ is below a certain threshold. Then given the size of this covariance matrix, we have to at least compute $\sim 2^N$ correlation functions, which is still the bottleneck exhibited in step 1.

With these heuristic discussions in mind, we summarize again with possible improvements:

| Step | Process | Complexity |
|---|---|---|
| 1 | Build covariance matrix $\Gamma$ | $2^{\mathcal{O}(N)}$ |
| 2 | Diagonalize $\Gamma$ | $J \cdot \frac{1}{\varepsilon} \text{poly}(N)$ |
| 3 | Compute entanglement entropy | $J$ |

The complexity summarized here, together with corollary 8 proves the result in Thm. 7.

On the CFT side, estimating the entanglement entropy of a quantum subregion to $\mathcal{O}(1)$ additive accuracy is computationally hard in general, even for physical states such as ground states of local Hamiltonians. The difficulty lies in the fact that the von Neumann entropy is a nonlinear, nonlocal functional of the reduced density matrix $\rho_A$, and capturing its value to constant precision requires access to its full spectrum. In practice, this entails either complete knowledge of $\rho_A$ or sufficiently precise approximations to its eigenvalues—both of which generally demand exponential resources in the size of the subregion.

From the complexity-theoretic perspective, this is the HQED problem, which is as hard as the QED and ED problems proved in [1], which remain intractable under standard cryptographic assumptions just as LWE. These results imply that, unless post-quantum cryptography is insecure, there can be no efficient algorithm (quantum or classical) that estimates the entropy of general quantum states—particularly mixed states prepared by shallow circuits or as ground states of local Hamiltonians—with additive error $\delta = \mathcal{O}(1)$. The hardness applies even when the state is physically realizable and highly structured, and even when multiple copies of the state are available.

In terms of methods, full quantum state tomography remains infeasible due to its exponential sample and computational complexity [32–34]. Approaches based on Rényi entropy estimation—such as swap tests or randomized measurements [26]—also exhibit exponential scaling in the number of required copies or circuit depth for high-rank states. Shadow tomography [35] and other low-rank approximation techniques similarly fail to provide entropy estimates with constant additive error in the worst case, unless additional structure or prior information is assumed.

Even for discretized CFT ground states, which in the continuum admit analytic entropy expressions, the problem persists. Once the state is encoded as a quantum circuit or tensor network over $N$ qubits, computing the entanglement entropy of a subregion to $\mathcal{O}(1)$ precision again requires reconstructing the full reduced density matrix, and thus faces the same exponential barriers. Small variations in high-rank spectra can lead to $\mathcal{O}(1)$ changes in entropy, making coarse-grained estimators unreliable unless one resolves fine-grained spectral structure.

For our example, in a $1+1$-dimensional conformal field theory (CFT), the entanglement entropy $S_A$ of an interval of length $\ell$ in the vacuum (with periodic boundary conditions and in the absence of thermal or excited-state contributions) takes the well-known universal form:

$$S_A = \frac{c}{3} \log\left(\frac{\ell}{\epsilon}\right) + S_{\text{subleading}}, \tag{116}$$

where $S_{\text{subleading}}$ encapsulates non-universal, finite corrections of order $\mathcal{O}(1)$ in a large-$c$ expansion.

The leading logarithmic term arises from global conformal symmetry and is universal across all unitary CFTs. In contrast, the subleading contributions $S_{\text{subleading}}$ encode more detailed, theory-specific data and stem from a number of distinct microscopic sources:

- Contributions of descendant states in the conformal block decomposition, corresponding to Virasoro blocks beyond the identity,

- Exchange of multi-stress-tensor operators in the operator product expansion (OPE) of twist operators,

- Nontrivial operator mixing in interacting theories, including contributions from marginal or irrelevant deformations,

- Finite-size effects and modular corrections arising from spatial curvature or finite volume geometries.

Importantly, these subleading terms are not determined by symmetry alone and depend intricately on the full operator content and correlation structure of the theory. From a bulk perspective, they correspond to quantum (loop-level) effects in the bulk entanglement wedge and reflect entanglement of bulk matter fields across the RT surface.

More formally, the task of estimating the von Neumann entropy of an unknown quantum state to within additive error $\delta = \mathcal{O}(1)$ is complete for the complexity class QSZK (Quantum Statistical Zero Knowledge) [36]. Therefore, distinguishing whether the entropy is greater or less by $\mathcal{O}(1)$ is believed to be computationally hard for quantum computers in general.

This reinforces the view that estimating entanglement entropy, even in highly structured quantum gravitational settings, is generically a computationally hard task. Building on the fact that measuring the leading-order term—the Ryu–Takayanagi formula—requires exponential complexity (see Thm.6), the subleading FLM correction, which involves the bulk entanglement entropy, is likewise not efficiently estimable, as established in Thm.7. On the boundary side, estimating the entropy of a reduced density matrix to $\mathcal{O}(1)$ additive accuracy also incurs exponential complexity, as supported by the hardness of the QED and ED problems under Learning with Errors (LWE)-based cryptographic

assumptions. Together, these results indicate that both bulk and boundary observers face a computationally intractable task when attempting to resolve fine-grained entropy, preserving the consistency of AdS/CFT with quantum computational limits.

We thus restate that both the bulk and boundary observers face a physical process—reconstructing entropy to constant accuracy—that is not efficiently simulatable on a quantum computer. As a result, Learning with Errors (LWE) remains secure even in the presence of holographic dualities, and the quantum-extended Church–Turing thesis (QECTT) is not violated: neither side of the AdS/CFT correspondence offers a computational shortcut that bypasses the exponential complexity of fine-grained entropy estimation. This computational equivalence across the duality further supports the consistency of holography with established quantum complexity-theoretic assumptions.

## IV. BENCHMARKING QUANTUM AND HOLOGRAPHIC ATTACKS ON LWE

It is also interesting to compare the best known quantum algorithm for attacking the LWE problem and our algorithm utilizing holographic duality.

A dominant classical approach to solving (search) LWE is to embed the samples into a lattice and then use lattice-reduction to recover the secret by decoding the small error vector [25, 37]. Formally, given $m$ samples $(\mathbf{A}, \mathbf{b} = \mathbf{A}\mathbf{s} + \mathbf{e} \mod q)$ one builds the usual LWE-embedding lattice $\mathcal{L}$ (determinant about $q^N$) so that solving LWE reduces to a closest-vector / bounded-distance decoding instance in $\mathcal{L}$. The analysis typically uses the Geometric Series Assumption (GSA): after a $\text{BKZ}_\beta$ reduction the shortest vector length behaves like

$$\lambda_1(\mathcal{L}) \approx \delta_\beta^N (\det \mathcal{L})^{1/N} , \tag{117}$$

where $\delta_\beta$ is the (empirical) root-Hermite factor delivered by blocksize $\beta$. Decoding succeeds when the error length $\|\mathbf{e}\| \approx \alpha q \sqrt{N}$ is smaller than $\lambda_1$, which (under the GSA) leads to the familiar scaling rule

$$\beta \simeq \frac{N \log q}{\log(1/\alpha)} . \tag{118}$$

For the common cryptographic regime $q = \text{poly}(N), \alpha = 1/\text{poly}(N)$ this gives $\beta = \mathcal{O}(N)$; since SVP in dimension $\beta$ is exponential, the overall runtime is exponential in $N$.

One promising idea is called the BKZ-based (primal/dual) attacks [37, 38]. Solving SVP in block dimension $\beta$ dominates the cost. Current best classical sieving algorithms give heuristics $T_{\text{SVP}}(\beta) \approx 2^{(0.292+o(1))\beta}$ (tuple-sieving and other engineering improvements change the constant and time/memory tradeoffs). Plugging $\beta = \mathcal{O}(N)$ yields an overall cost roughly $2^{cN}$ for some small constant $c$ (concrete estimators report $c$ in the low $-10^{-3}$ to $10^{-2}$ range for typical parameters).

This can be improved with quantum lattice sieving speedups [37, 38]. Quantum algorithms reduce the asymptotic sieve constant modestly (heuristically to about $2^{(0.265+o(1))\beta}$ for pure quantum sieving) and give polynomial or modest exponential improvements for the SVP subroutine - but they do not yield a polynomial-time break. Concrete analyses show only a modest security margin reduction once realistic quantum overheads and memory are considered.

Another approach is called the dual / decoding-based attacks [39]. The dual attack builds short vectors in the dual lattice to correlate with the error. For small or low-entropy secrets (binary/ternary or sparse), the distinguishing step becomes easier and dual/hybrid attacks can substantially lower the effective exponent relative to the primal attack. Recent work has tightened and extended these dual techniques (including improved distinguishers and decoding steps).

The BKW-style algorithms and their modern enhancements (coded-BKW, lazy modulus switching, FFT distinguishers, BKW+ sieving) are subexponential in the standard LWE dimension [40, 41]:

$$T_{\text{BKW}}(N) \approx 2^{\mathcal{O}(N/\log N)} , \tag{119}$$

and they become competitive when the noise is very small, modulus is small, or when a huge number of samples is available. Coded-BKW and other engineering tricks have pushed practical costs down in those regimes.

Combining partial lattice reduction with combinatorial search or guessing (e.g., reduce dimension with BKZ then run BKW / meet-in-the-middle on remaining coordinates) often yields the best practical tradeoffs against special secret distributions (sparse or structured secrets) [42].

Comparing Thm. 7 with the best-known algorithms for attacking LWE shows that the holographic reconstruction problem and standard LWE attacks lie in the same asymptotic hardness class: both require resources that grow exponentially in the number of degrees of freedom $N$. Concretely, your theorem predicts a measurement cost

$$\mathcal{T}(N) = \text{poly}(N) e^{N^{1/\alpha}} 2^{\mathcal{O}(N)} = \exp(\mathcal{O}(N)) , \tag{120}$$

for $1 \lesssim \alpha$, and standard lattice-reduction attacks (BKZ + SVP/sieving) yield runtimes of the form $2^{\mathcal{O}(N)} = \exp(\mathcal{O}(N))$ as well. Thus, in the coarse asymptotic sense both barriers are exponential in $N$. It is important to stress, however, that this coincidence of complexity classes masks two critical practical distinctions: (i) the leading constant in the exponential matters numerically for feasible $N$ and can make one task far more costly than the other in practice, and (ii) the resource semantics differ - LWE hardness is primarily a computational/time/memory barrier, whereas the holographic bound is an information/measurement barrier with distinct operational constraints and specific situations we encounter for our bulk fields (dominant symplectic eigenvalues, sparsity of the covariance matrix, etc.) We therefore recommend reporting both the asymptotic equivalence and a numerical comparison of exponent constants for representative $N$ when arguing about "practical" intractability.


}

\end{document}